\documentclass[10pt]{iopart}

\usepackage[utf8x]{inputenc}
\usepackage{graphicx}
\usepackage{caption}
\usepackage{subcaption}
\usepackage{float}
\newtheorem{theorem}{Theorem}
\usepackage{url}
\usepackage[colorlinks=true, allcolors=blue]{hyperref}
\usepackage{fancyhdr}
 \usepackage[colorinlistoftodos]{todonotes}

\begin{document}
\title[`Sandwich' meta-framework for deep PP transfer learning]{The `Sandwich' meta-framework for architecture agnostic deep privacy-preserving transfer learning for non-invasive brainwave decoding}

\author{Xiaoxi Wei$^1$, Jyotindra Narayan$^{1,2}$, A. Aldo Faisal $^{1,2}$}

\address{$^{1}$ \quad  Brain \& Behaviour Lab, Department of Computing, Imperial College London, London SW7 2AZ, United Kingdom.\\
$^{2}$ \quad  Institute of Artificial \& Human Intelligence, University of Bayreuth, Bayreuth 95447, Germany}



\begin{abstract}
\emph{Objective}. Machine learning has enhanced the performance of decoding signals indicating human behaviour. EEG brainwave decoding, as an exemplar indicating neural activity and human thoughts non-invasively, has been helpful in neural activity analysis and aiding paralysed patients via brain-computer interfaces. However, training machine learning algorithms on EEG encounters two primary challenges: variability across data sets and privacy concerns using data from individuals and data centres. Our objective is to address these challenges by integrating transfer learning for data variability and federated learning for data privacy into a unified approach.
\emph{Approach}. We introduce the `Sandwich' as a novel deep privacy-preserving meta-framework combining transfer learning and federated learning. The `Sandwich' framework comprises three components: federated networks (first layers) that handle data set differences at the input level, a shared network (middle layer) learning common rules and applying transfer learning techniques, and individual classifiers (final layers) for specific brain tasks of each data set. This structure enables the central network (central server) to benefit from multiple data sets, while local branches (local servers) maintain data and label privacy.
\emph{Main Results}. We evaluated the `Sandwich' meta-architecture in various configurations using the BEETL motor imagery challenge, a benchmark for heterogeneous EEG data sets. Compared with baseline models like \emph{Shallow ConvNet} and \emph{EEGInception}, our `Sandwich' implementations showed superior performance. The best-performing model, the Inception SanDwich with deep set alignment (\emph{Inception-SD-Deepset}), exceeded baseline methods by 9\%.
\emph{Significance}. The `Sandwich' framework demonstrates advancements in federated deep transfer learning for diverse tasks and data sets. It outperforms conventional deep learning methods, showcasing the potential for effective use of larger, heterogeneous data sets with enhanced privacy. In addition, through its diverse implementations with various backbone architectures and transfer learning approaches, the `Sandwich' framework shows the potential as a model-agnostic meta-framework for decoding time series data like EEG, suggesting a direction towards large-scale brainwave decoding by combining deep transfer learning with privacy-preserving federated learning.

\end{abstract}

\noindent{\it Keywords}: Deep Learning, Transfer Learning, Domain Adaptation, Brain-Computer-Interfaces (BCI), Electroencephalography (EEG), Privacy-preserving, Federated Machine Learning, Heterogenous Datasets


\maketitle


\section{Introduction}

Disabilities, paralysis, and conditions such as stroke exert a profound impact on individuals, significantly impairing their ability to engage with their environment. Brain-computer interfaces (BCIs) have emerged as a promising solution aimed at assisting individuals in regaining lost functionalities \cite{vaid2015eeg,rashid2020current}. BCIs represent interfaces designed to decode the intricate patterns of human brainwave activity and translate them into computer commands. These commands are subsequently employed to oversee real-time systems, e.g. wheelchairs or video games. Advancements have been achieved in the field of invasive BCI methodologies\cite{ramadan2017brain}. However, invasive BCI typically necessitates surgical interventions, which are not universally accessible or desirable. Among the various non-invasive BCI techniques, Electroencephalography (EEG) stands out as a safe and relatively non-expansive choice. EEG entails the measurement of cortical field potentials emanating from the scalp's surface. Within this context, the application of deep learning has emerged as a potent approach, obviating the need for manual feature engineering while eliminating inherent biases \cite{walker2015deep, BallCNN}. Nevertheless, EEG recordings often exhibit challenges of low signal-to-noise ratio and cross-subject or cross-dataset variabilities stemming from differences in scalp sensor placements, sensor impedance, individual brain anatomy, functional organisation and different tasks performed during data collection. Consequently, BCI systems typically mandate an extensive training period, sometimes for months, for new users to attain a personalised and stable decoding mechanism, thereby impeding the seamless integration of BCIs into real-world applications. To ameliorate these challenges, our study employs a combination of deep learning and transfer learning methodologies to enhance EEG-BCI decoding.

Deep learning has been widely used for EEG decoding \cite{vaid2015eeg,rashid2020current}. Convolutional neural networks (CNNs) have gained widespread adoption in the domain of EEG decoding across various fields, including applications like seizure detection, Parkinson's disease diagnosis, and motor imagery decoding, as evidenced by prior studies \cite{4,5,6}. CNNs have risen to prominence and serve as benchmarks in fields like motor imagery decoding in the existing literature \cite{3}. CNNs offer a distinct advantage in their ability to extract relevant features from raw EEG data autonomously. This autonomous feature learning capability is especially valuable as it obviates the need for subjective decisions often made during handcrafted EEG processing, thereby mitigating the risk of biased feature extraction \cite{23,24}. 

Nonetheless, the utilisation of deep learning techniques in EEG analysis poses a notable challenge known as the negative transfer problem, as highlighted in previous research \cite{25,wei2021inter}. This challenge manifests as a significant drop in the accuracy of CNNs when they are trained on EEG data sourced from multiple subjects or data sets, which is caused by the disparate distributions across data sets. The underlying source of this dissimilarity stems from inter-subject variability, encompassing differences in skull shapes, brain anatomies, and brain functionalities. Consequently, due to this negative transfer effect, it becomes impractical to directly apply EEG data from other individuals, data sets and tasks to compensate for the scarcity of relevant data. Moreover, CNNs are renowned for their requirement for large amounts of data. Consequently, most BCI solutions continue to necessitate the collection of extensive EEG data sets from users across numerous sessions to attain stable and desirable levels of accuracy. These prolonged training periods impede user motivation and stand as one of the principal challenges confronting the field of BCI. 

Transfer learning has emerged as a promising solution to effectively merge data from various data sets, even those originating from domains with distinct data distributions. Transfer learning \cite{weiss2016survey} encompasses a wide array of algorithms with the overarching goal of transferring knowledge and representations gleaned from source domains to a target domain. Some strategies have been developed in this context, primarily within fields such as computer vision and natural language processing. One straightforward approach is fine-tuning, frequently employed to transfer model representations from one specific task or data set to another \cite{11}. In fine-tuning, neural networks are initially pre-trained on one domain and subsequently fine-tuned on the target domain, effectively adapting their learned features to the new data distribution. Another strategy involves training a shared network in early layers that simultaneously incorporates all available data sets but distinguishes and adapts the deeper layers to accommodate the unique characteristics of each data set \cite{12}. Moreover, deep domain adaptation \cite{13} represents a technique that directly maps source and target data distributions into a unified distribution using Neural Networks. With this approach, methods enable data from source domains to contribute to improving the model's performance on the target domain, bridging the gap between different data distributions.

Some studies have tried to apply these concepts to EEG decoding\cite{jayaram2016transfer, lotte2018review}. In one notable example \cite{14}, a deep transfer network was introduced to transfer knowledge representations from images to EEG images. Another study enhanced domain adaptation by incorporating an adversarial network into the architecture, leading to improved performance \cite{7}. Additionally, in the realm of motor imagery decoding, pre-training and fine-tuning strategies were leveraged in a study \cite{15} to develop effective models. Moreover, a deep domain adaptation network was utilised to facilitate the transfer of data representations between different subjects in EEG decoding \cite{17}, with a focus on separating the deeper layers of the network to accommodate the individual subject differences. However, despite these advancements, two challenges remain in the literature. Firstly, while current domain adaptation networks are effective in transferring data distribution from one data set to another (one-to-one transfer), successfully addressing the transfer of multiple data sets to a single target (multiple-to-one transfer) has not been fully resolved. Secondly, the field still grapples with the issue of transfer learning involving multiple EEG decoding tasks, which has yet to find a comprehensive solution in the literature. These challenges constrain the ability to leverage a diverse range of data for training deep-learning models with the highest possible accuracy.

A significant attempt to address these challenges was the international BEETL BCI competition, which took place at NeurIPS 2021 \cite{pmlr-v176-wei22a}. The competition centred around cross-dataset and cross-task EEG transfer learning and garnered substantial academic attention. Various successful solution algorithms using deep transfer learning were developed to tackle the BEETL challenge, offering insights for cross-dataset and cross-task EEG transfer learning. Some other studies in the realm of EEG transfer learning have explored various distribution alignment approaches to align data sets and domains from different sources. One of the techniques includes Maximum-Mean Discrepancy (MMD), which has been employed for inter-subject transfer learning in EEG decoding \cite{smola2006maximum,13, wei2021inter}. Maximum Density Divergence (MDD) has also been utilised as a metric for measuring distributional dissimilarity in EEG data transfer scenarios \cite{li2020maximum,han2023eeg}. Moreover, set theory combined with deep learning was introduced into EEG transfer learning \cite{zaheer2017deep,pmlr-v176-wei22a,bakas2022team} to deal with variability across data sets. 

However, another challenge arises for cross-dataset transfer learning: the imperative of privacy preservation. As machine learning algorithms increasingly demonstrate their capacity to harness heterogeneous EEG data sets sourced from various data centres and providers for large-scale applications, the safeguarding of EEG data privacy becomes a primary concern. It is crucial to acknowledge that brainwave data contains intricate and sensitive information, encompassing images, text, and even individual identities, all of which machine learning algorithms possess the potential to unveil. Consequently, the practice of sharing EEG data across data centres or international borders encounters substantial regulatory and legal restrictions, exemplified by stringent frameworks such as the EU's GDPR data policy and ongoing data disputes involving nations like China and the United States.

There are some strategies for privacy-preserving in the machine learning literature \cite{li2020review, hao2019towards, lyu2017privacy, dong2017dropping, du2001secure}. Federated Learning \cite{li2020review,truong2021privacy} stands as an approach where models are trained on edge servers, eliminating the need for raw data exchange. While this method effectively safeguards privacy, it maintains the ability for collaborative model training, striking a balance between data security and model improvement. Data Encryption Methods \cite{hao2019towards} take the path of encrypting either the raw data or model parameters, yet this approach introduces an encoding-decoding procedure that comes at the cost of additional computational resources. Data Transformation Techniques \cite{lyu2017privacy}, such as the introduction of cancelable noise to local gradients or parameters prior to their transmission to a central server, offer a means of safeguarding individual data privacy, which also introduces extra computational costs. These techniques obscure sensitive information while allowing for secure data sharing. Model Splitting Methods \cite{dong2017dropping} allocate distinct model parameters to various data sets, ensuring the protection of each individual's model privacy. This approach contributes to enhanced data security within collaborative settings. Multi-party Computation \cite{du2001secure} involves local model training by individual parties followed by secure aggregation of model updates conducted by a trusted third party. This method protects privacy while enabling the improvement of the collaborative model.

Nonetheless, the suitability of these methods for brainwave decoding has not been universally established. Furthermore, these approaches may introduce additional computational costs, which could impede training efficiency, particularly when dealing with large-scale EEG data sets. Balancing privacy with the practicalities of EEG data analysis remains an ongoing challenge in brain-computer interfaces. Several studies within the EEG literature have delved into privacy preservation \cite{popescu2021privacy, xia2022privacy, ju2020federated, bethge2022domain}, leveraging the methods previously discussed. Additionally, there have been endeavours to enhance privacy preservation through improved protocols and user-level system design \cite{kapitonova2022framework}. Recent investigations \cite{wei2021inter, bethge2022domain} explored the deployment of distributed individual feature extractors, allowing for the accommodation of EEG differences of individual data sets while upholding data privacy without incurring the additional costs associated with encryption. However, it's important to note that the studies mentioned thus far could not fully address the challenge of cross-dataset transfer learning when handling different tasks while simultaneously maintaining the privacy of data, parameters and inference. This limitation can restrict the utilisation of large-scale EEG data in real life. Consequently, the development of privacy-preserving methods capable of accommodating diverse tasks within cross-dataset transfer learning remains an ongoing challenge in the EEG research landscape.

In summary, below we highlight three challenges in the literature that pose potential barriers to the advancement of large-scale EEG transfer learning. Firstly, addressing the variability in experiment setups in different data sets, including differences in EEG input shapes, impedance, and sensor positions, is still a challenge. A solution is needed to harmonize these disparities for effective cross-dataset transfer learning. A second challenge is learning common knowledge across data sets as well as being able to differentiate the information of each individual data set. The key challenge is to find a solution that learns common knowledge across data sets and maintains their individual properties. Developing approaches that enable the incorporation of diverse tasks into a unified model is pivotal for comprehensive EEG transfer learning. The final challenge is to build a transfer learning model that integrates privacy-preserving properties. Ensuring the privacy of EEG data stored in various data centres is important for encouraging secure and seamless data sharing. Robust methods for privacy preservation should be integrated into the transfer learning process to facilitate collaborative research while safeguarding sensitive information. These challenges collectively underscore the complexity of large-scale EEG transfer learning and emphasise the need for innovative solutions that address these issues to unlock the full potential of EEG data in diverse applications.

\begin{figure*}
\centering
\includegraphics[scale=0.4]{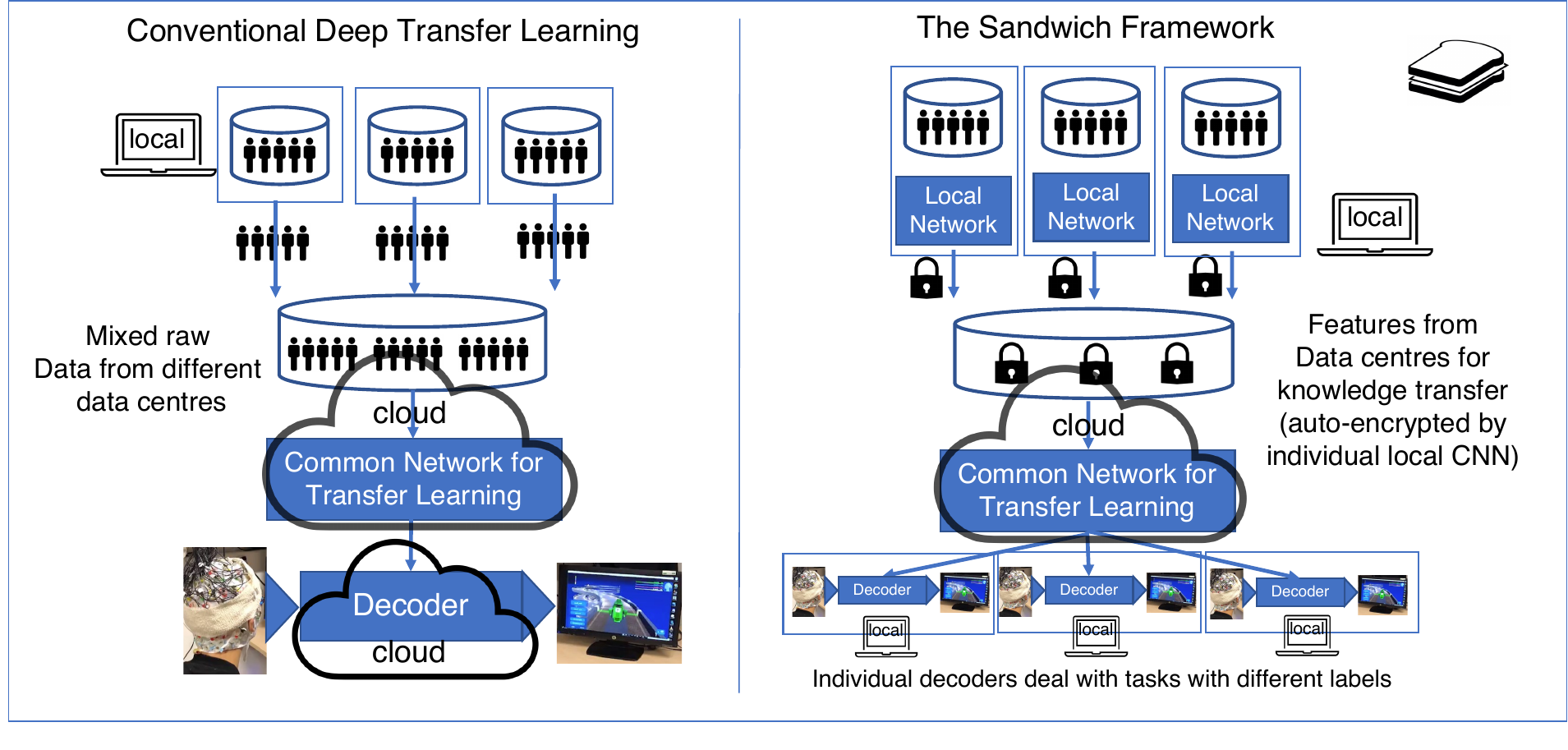}
\caption{Demonstration of the privacy-preserving deep transfer learning. Conventional transfer learning involves pooling source data sets on a central server, whereas, in the Sandwich framework, data are encrypted as features in the local network before common knowledge training. Moreover, the privacy-preserving transfer learning framework enables predictions to be made on local branches, ensuring data privacy and security from the prediction level.}.
\label{concept}
\end{figure*}

In our research, we introduce a model-agnostic deep privacy-preserving meta-framework to tackle the aforementioned challenges, termed the `Sandwich' due to its 3-layer structure, a demonstration of the concept is shown in Figure \ref{concept}. The Sandwich framework is specifically engineered to integrate privacy-preserving features into deep transfer learning for EEG decoding across multiple data sets and tasks. Our study presents various implementations of the meta-architecture to show its versatility, as well as demonstrating improvements in decoding accuracies on the BEETL BCI challenge when compared to baseline methods. Moreover, we introduce the enhanced privacy-preserving capabilities and the ability to utilise heterogeneous data sets, underscoring its potential to advance the field of brain-computer interfaces towards larger-scale EEG transfer learning.
\section{Methods}

In this section, we will introduce our Sandwich meta-architecture, followed by the presentation of two exemplary implementation architectures, \emph{Shallow-ConvNet-Sandwich} and \emph{EEGInception-Sandwich}, on the BEETL BCI challenge. These examples will demonstrate how to seamlessly integrate existing EEG decoding models into the federated transfer learning meta-framework. Subsequently, we will delve into the demonstration of two transfer learning approaches employed within the Sandwich framework, the distributional alignment and deep set approach.

\begin{figure*}
\centering
\includegraphics[scale=0.45]{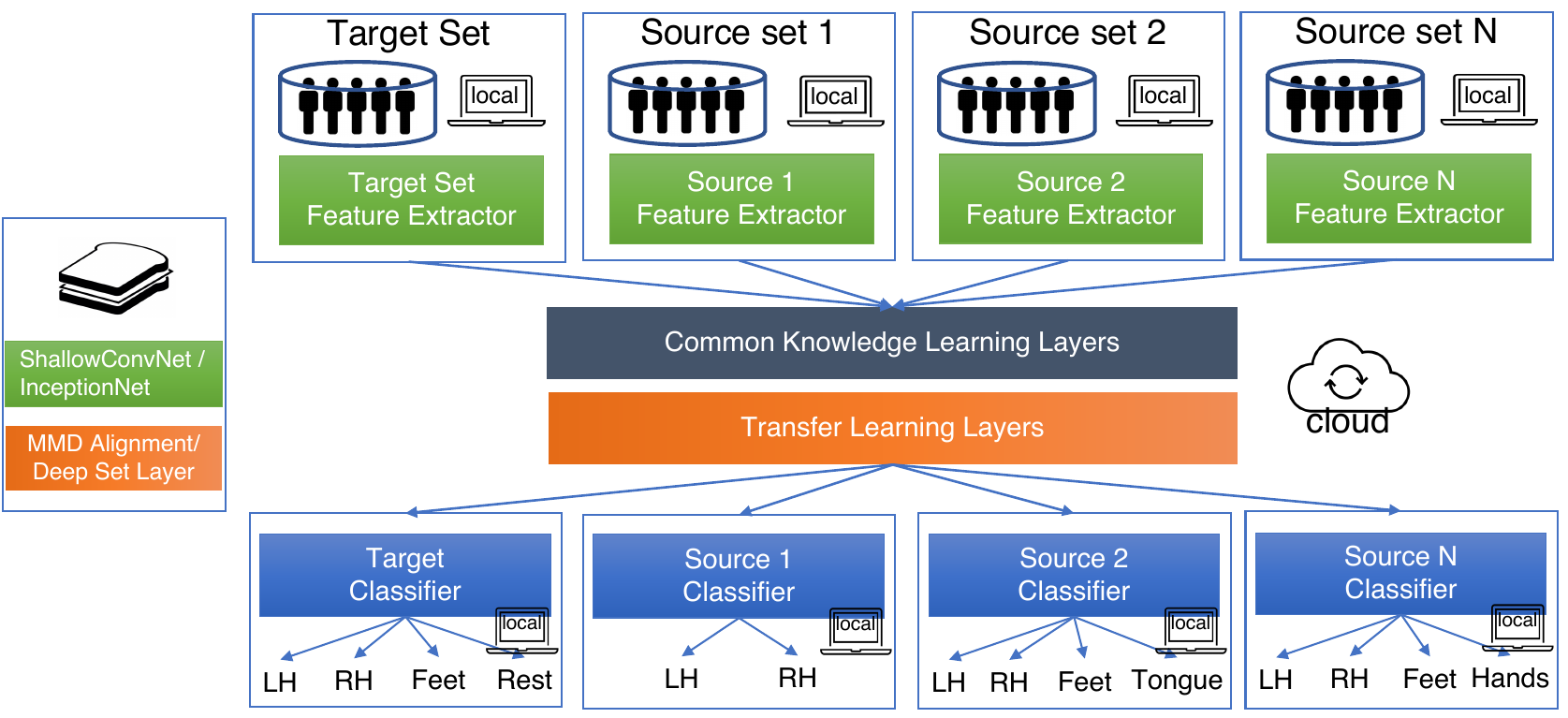}
\caption{The network architecture of the Sandwich with multiple classifiers}.
\label{SCSNDeepset}
\end{figure*}

\begin{figure*}
\centering
\includegraphics[scale=0.45]{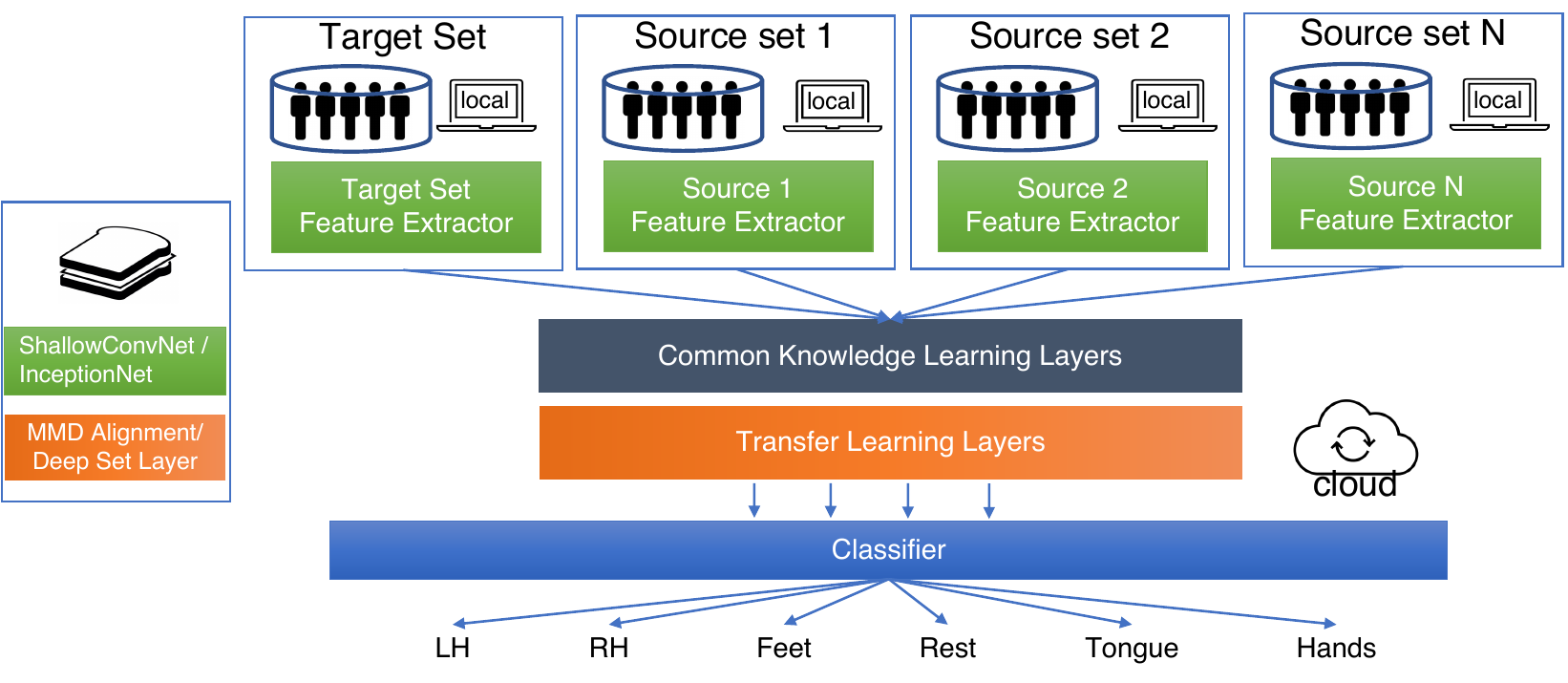}
\caption{The network architecture of the Sandwich with a unified classifier}.
\label{SCSNMMD}
\end{figure*}

The general framework of the Sandwich is illustrated in Figure \ref{SCSNDeepset} and \ref{SCSNMMD}. It begins with individual split branches functioning as feature extractors for each data set. These individual feature extractors serve two primary purposes. Firstly, they handle discrepancies in the original data setup, including variations in channel impedance, location and number of channels. Across subjects or datasets, even when the data sets share the same channel names, locations could be shifted, and impedance, as well as channel sizes, vary. Secondly, these individual feature extractors can be locally stored and trained within each data centre, effectively serving as auto-encryption layers alongside feature extraction. This process can be likened to an auto-encryption mechanism, where the individual feature extractors are exclusively utilised within their local server during training, ensuring that the `key' to encryption remains solely in the possession of the local server. Only the features `encrypted' are shared with the central server for common knowledge learning and transfer learning. In this way, the framework ensures raw-data level and parameter-level privacy to be secured.

Following the individual feature extractors are the common layers. The functionality of the common layers is to learn common knowledge across different data sets. These common layers receive features from each branch and perform a common feature extraction process, facilitating shared rules for feature extraction across different data sets. After common knowledge learning, additional deep transfer learning techniques can be applied within the Sandwich framework, such as distributional alignment \cite{smola2006maximum} or the deep set approach \cite{pmlr-v176-wei22a}. These transfer learning layers could be integrated into or after the common knowledge transfer learning layers. A detailed description of network parameters of implementation and computations regarding these additional transfer learning layers will be presented later in this section.

Two strategies for the final classification are considered to address variations in labels. The first strategy involves using a unified classifier that encompasses all tasks from all data sets (Figure \ref{SCSNMMD}). This approach potentially benefits the classifier by exposing it to a broader range of training examples, particularly if these examples are drawn from similar distributions after the alignment. The second strategy entails splitting the classifiers for different data sets, which may perform better when tasks exhibit very dissimilar distributions (Figure \ref{SCSNDeepset}). Additionally, the second strategy introduces an additional layer of privacy preservation, as these classifiers can be trained and make predictions on local servers, which protects model inference-level privacy.

To illustrate the implementation of the Sandwich meta-framework, we incorporate two example models into the architecture. These models serve as tangible demonstrations of how the Sandwich can be effectively utilised. For establishing a comparison baseline, we employ the Shallow ConvNet \cite{3} and EEG InceptionNet \cite{santamaria2020eeg}, two benchmark models for motor imagery BCI decoding, as baseline models. 

We first introduce the implementation of the Sandwich framework using the EEG \emph{InceptionNet}. We employ a feature extractor design similar to that utilised in the BEETL winning method, in which the detailed network design can be found \cite{bakas2022team, pmlr-v176-wei22a}. The main concept of the inception network layer involves channelling the data through multiple parallel layers with distinct kernel sizes. Subsequently, the filtered features from these parallel layers are concatenated before being forwarded to the next layer. Specifically, the individual inception feature extractor used in this study comprises a temporal inception block composed of three inceptions with kernel sizes of (1, 21) but varying dilation rates of (1, 4), (1, 2), and (1, 1). Then, a spatial convolutional layer with a kernel size of (channel size, 1), and a filter size of 48 is applied. Additional components afterwards include a batch normalisation layer, an average pooling layer, and a dropout layer with a dropout rate of 0.25. The common knowledge learning layer consists of another inception block with a kernel size of (1,5) and varying dilation rates of (1, 8), (1, 4), and (1, 2). The transfer learning layers consist of two CNN blocks as learning parameters. The first CNN block has two convolutional layers of kernel size (1, 9) and (1,1) with a batch normalisation layer and a dropout of 0.25. The other CNN block has smaller kernel sizes of (1, 5) and (1,1) with a batch normalisation layer and a dropout of 0.25. In this study, two transfer learning approaches were tested in the transfer learning layers. If MMD alignment is applied, an MMD layer is conducted once after the second CNN block. If the deep set method is used, the deep set layers are applied both before and after the CNN blocks. We will introduce how MMD and Deep Set work in later paragraphs. Finally, a standard unified 6-class classification layer or multiple individual classifiers are used to perform classification.

Furthermore, we integrate the \emph{Shallow ConvNet} as a second example of the feature extractor for the Sandwich implementation.  This additional assessment demonstrates the versatility of Sandwich in accommodating various EEG decoding models. Similar to the Shallow ConvNet study \cite{3}, the Shallow ConvNet feature extractor comprises several key components: a temporal layer with a kernel size of (25,1) and 40 filters, a spatial layer with a kernel size of (1, channel size) and 40 filters, a batch normalization layer, a square non-linear layer, a pooling layer with a kernel size of (75,1) and a stride size of (15,1), a logarithmic non-linear layer for pooling, and finally, a dropout layer with a rate of 0.5. The above layers assume data format as (time samples, and channels). These architectural parameters remain consistent and can be found in the study of Shallow ConvNet \cite{3}. The temporal layer is responsible for extracting time-relevant information, while the spatial layer performs spatial convolutions across EEG channels, collectively contributing to the feature extraction process. Before the common transfer layers, a feature reduction layer was applied to reduce the feature space to 50. The common knowledge learning layer consists of three convolutional layers with a filter size of 50 and kernel size of (1,1). If transfer learning layers are applied, an MMD layer is conducted after the three convolutional layers. The deep set layers are added both before and after the convolutional layers if the deep set method is used. Finally, the network has a standard unified 6-class classification layer or multiple individual classifiers to perform prediction.

Following the introduction of the network structures, we present the two transfer learning techniques in the subsequent paragraphs: domain adaptation based on maximum-mean discrepancy (MMD) \cite{smola2006maximum} for aligning distributions and deep set alignment \cite{pmlr-v176-wei22a} to enhance set invariance. These two transfer learning techniques are applied to the networks to boost transferability across different data sets.

In this study, we have devised a Sandwich-MMD architecture, incorporating Maximum Mean Discrepancy (MMD) alignment into the Sandwich framework. Long et al. \cite{long2015learning} have successfully employed MMD as a similarity constraint to reduce the distributional divergence between just two image domains in Deep Domain Adaptation Networks (DAN). In our study, we extend the DAN framework to our proposed network architectures by incorporating Maximum Mean Discrepancy (MMD) computation among multiple EEG data sets. This adaptation enhances our network's ability to align distributions across diverse data sets, contributing to improved transfer learning capabilities. The detailed computation of MMD is elucidated in Equation \ref{Equation1}. $X_t$ and $X_{si}$ represent the target data and the $i_{th}$ source data respectively. The selection of the kernel function $\Phi$ has been a prominent subject of investigation in the existing literature, and in line with prior research \cite{long2015learning}, we employ the Radial Basis Function (RBF) kernel with the mean L2 distance of the data serving as the variance. MMD is then included as a component of the loss function. In the loss function, denoted by Equation \ref{Equation2}, $L_c$ signifies the classification loss, N represents the count of source data sets, and $\lambda$ serves as the weight parameter that governs the trade-off between the MMD loss and the classification loss. Within the Sandwich-MMD framework, we compute and accumulate $MMD_i^2$ between the output features of the target branch network and each of the source branch networks $i$. Our study concentrates on a common task across all data sets, specifically left-hand and right-hand motor imagery. To this end, our model computes the MMD distance with respect to left-hand and right-hand samples between the source and target domains.

\begin{eqnarray}
\mbox{MMD}_i^2(X_{si},X_t) = ||\frac{1}{{X_s}_i}\sum_{{x_s}_i\in {X_s}_i}\Phi ({x_s}_i)  - \frac{1}{X_t}\sum_{x_t\in X_t}\Phi (x_t)||.
\label{Equation1}
\end{eqnarray}

To be specific about the implementation of the MMD layer mentioned in the architecture. It comprises a convolutional feature reduction layer, reducing dimensionality from the previous layer to 50, and another convolutional layer with 50 filters and a kernel size of (1,1). These layers act as learnable parameters, facilitating the projection of features into a comparable distribution. Subsequently, MMD distances, serving as distance losses, are computed based on Equation \ref{Equation2} to align the distribution of the target data set with each source data set.


\begin{eqnarray}
L = L_c + \lambda \sum_{i=1}^{N} (\mbox{MMD}_i^2(X_{si}^L,X_t^L)+\mbox{MMD}_i^2(X_{si}^R,X_t^R)).
\label{Equation2}
\end{eqnarray}

The deep set alignment layer is implemented based on the Deep Set \cite{zaheer2017deep}. Deep Set is a technique that employs deep learning to process sets of variable-size inputs. It addresses the challenge of working with unordered and varying sets using permutation-invariant architectures to align these source sets. Deep Set has applications in various domains and enables tasks such as set classification, matching, generation, and similarity estimation. The deep set idea was proposed to be used in EEG decoding, and we employed the identical configuration for the deep set layer as outlined in \cite{pmlr-v176-wei22a, bakas2022team,bakas2023latent}. The design of the deep set layers is grounded in Theorem \ref{TheoremDeepset}, in which each subject from different data sets is regarded as a `set'. In deep learning, convolutional layers can be interpreted as the transformation function $\Phi$, while the non-linear activation functions of these layers are denoted as nonlinearity $\rho$. 

\begin{theorem}[Set Theorem used in Deep Set \cite{zaheer2017deep}]
Function f(X) operating on a set X having elements from a countable universe, is a valid set function, i.e., invariant to the permutation of instances in X, iff it can be decomposed in the form $\rho(\sum_{x\in X} \Phi (x))$, for suitable transformations $\Phi$ and $\rho$.
\label{TheoremDeepset}
\end{theorem}

In alignment with the findings from previous studies, the deep set layers serve as a mechanism for alignment, rendering each set invariant to trial orders and enhancing the robustness of predictions within each set. A comprehensive description of the deep set layer implementation can be found in \cite{pmlr-v176-wei22a, bakas2022team,bakas2023latent}. In these studies, EEG trials underwent processing through convolutional layers. Subsequently, the EEG features of each subject are organised as `sets' and are then subjected to deep set layers to enhance the robustness of individual features. In our study, as illustrated in Figure \ref{SCSNDeepset} and \ref{SCSNMMD}, we further extend this concept to multi-task scenarios. The common layers facilitate knowledge transfer learning across data sets, and the subsequent deep set layers treat each subject as a set, reinforcing the stability and robustness of these features.

In this study, we employed the same deep set alignment block as the above studies. The deep set block rearranges the three-dimensional training batch from (S*T, N, f) to subject-wise four-dimensional features (S, T, N, f). Here, S represents subjects, T represents subject trials, N represents filter size, and f represents feature size. The mean of each subject is computed, resulting in a feature space of (S, 1, N, f). These subject-specific features are then passed through linear layers of size 8 resulting in a feature vector of (S, 1, 8, f). This process aligns with the concept of the `$\sum$' operation in Theorem \ref{TheoremDeepset}. Subsequently, these subject-specific features are concatenated back to each trial of the corresponding subject, serving as `global' information for that subject (S, T, N+8,f). The data is then passed through another linear layer to project the feature space back to the original size (S, T, N,f). This process corresponds to the $\Phi$ operation in Theorem \ref{TheoremDeepset}, and the activation function of the layers is regarded as the non-linear operation $\rho$ in the theorem. 

\section{Dataset and Experimental Protocol}

\begin{figure*}
\centering
\includegraphics[scale=0.45]{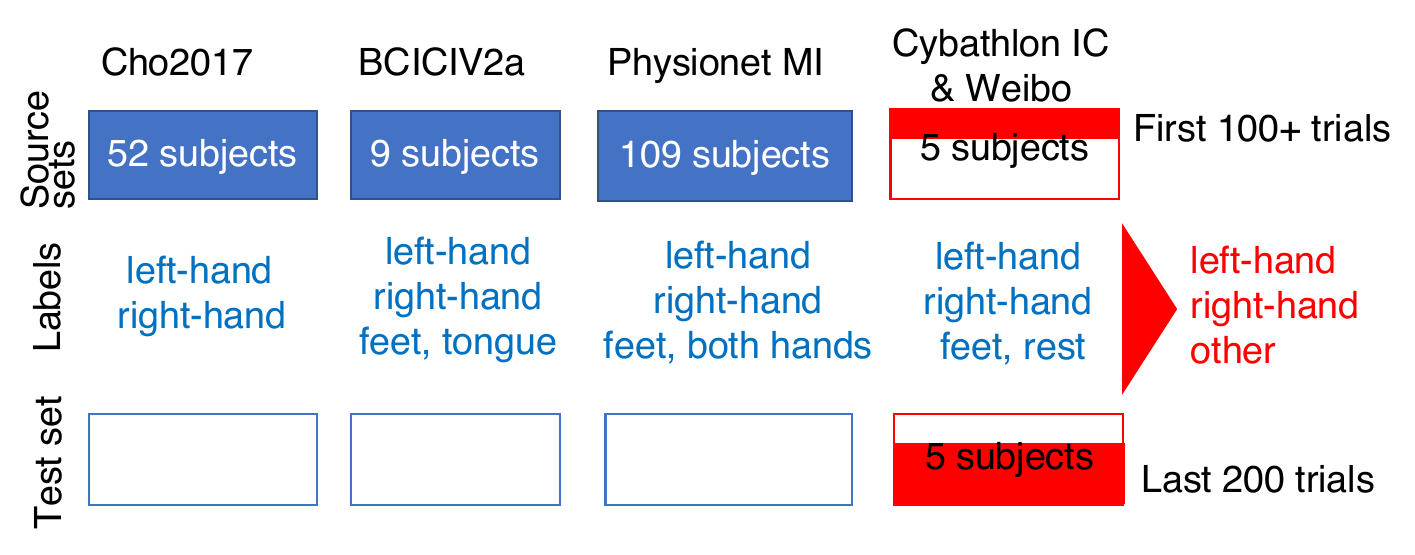}
\caption{The source data sets (columns 1-3 in blue) and the test set (column 4 in red) used in the BEETL BCI challenge. All data from the source sets (as the source domain) and the first 100 trials from the testing subjects (as the target domain) are used as the training set. The test set comprises the final 200 trials of each test subject. The top row of rectangles identifies the size of each data set. The middle row (labels) identifies the differing motor imagery labels used in each data set. The bottom row identifies the test set.}
\label{taskDemo}
\end{figure*}

The proposed architecture underwent testing using the motor imagery BCI task of the BEETL competition as an example of cross-dataset and cross-task real-world challenges. Comprehensive details regarding the dataset can be accessed in \cite{pmlr-v176-wei22a} and \url{https://github.com/XiaoxiWei/NeurIPS_BEETL}. To provide a visual overview, Figure \ref{taskDemo} and Table \ref{MIdataset} depict the data sets employed in the BEETL motor imagery (MI) task. The dataset comprises three distinct source data sets, each characterised by unique devices and data collection protocols. Specifically, the Cho2017 data set~\cite{Cho2017} encompasses 52 subjects engaged in left and right-hand motor imagery tasks. Moving on, the BCICIV2a dataset~\cite{BCIC} involves nine subjects participating in motor imagery tasks involving left-hand, right-hand, feet, and tongue actions. We selected subjects 1, 3, 7, 8, and 9 based on superior data quality, narrowly defined as embedding better discriminative features, as our sources. The selection is for a better quality of transfer in case of transferring noise from the source. Data quality per subject was determined from classification performance (independent of our work), as they were reported by previous studies or our replications of those studies \cite{hwaidi2022classification,schalk2004bci2000, wei2021inter}. The PhysioMI dataset~\cite{schalk2004bci2000,11}, the third source dataset, boasts a substantial 109 subjects. From this extensive pool, we chose subjects 1, 7, 17, 24, 28, 31, 33, 34, 35, 42, 49, 52, 54, 55, 56, 60, 62, 63, 68, 71, 72, 73, 85, 91, 93, 94, and 103 as our sources. 

The two baseline models and implementations of the Sandwich framework were evaluated using the BEETL BCI Challenge. In the test set, five subjects from two different data sets were included in the MI challenge, each contributing 200 trials (amounting to a total of 1000 trials with each category comprising 250 trials). The first three subjects (S1, S2, S3) were obtained from the CybathlonIC data set \cite{wei2021inter,pmlr-v176-wei22a}, which was collected through an online closed-loop format with a virtual car game, labelled as Rest (0), Left-hand (1), Right-hand (2), Feet (3). The remaining two subjects were sourced from the Weibo2014 data set \cite{weibo}, labelled as Left-hand (0), Right-hand (1), Feet (2), and Rest(3). The channel positions of test set A and B are shown in Figure \ref{ChannelPositions} left and right respectively. Common channels used for all testing and training sets are shown in the middle. It is important to note that, during training and prediction of models, it remains a 4-class classification problem. During scoring, as in the BEETL competition, we evaluate the final accuracies based on the weighted accuracy of three key classes by combining labels the last two labels as 'others'. In the BEETL MI challenge, to increase the variability of the test set, `other' in set A represents labels 2(RH) and 3(Feet), whereas 2(Feet) and 3(Rest) in set B. This design of challenge increases the variety of the two data sets and requires the algorithm to be generalised well to different task scenarios. In this study, we train set A (S1-S3) and set B (S4 and S5) individually as two models during training and testing.


To ensure the alignment and effective utilisation of all data sets, 3-second windows were extracted for all trials, given that Cho2017 imposed a maximum trial length of 3 seconds. It's worth noting that this may not fully exploit the potential of the 4-second target test data, as observed in other solutions excluding Cho2017 within the competition. Another comparison experiment to the winning method in the competition will be conducted later with a 4-second window. From the four data sets, we selected 17 common EEG channels (Fz, FC1, FC2, C5, C3, C1, C2, C4, C6, CP3, CP1, CPz, CP2, CP4, P1, Pz, P2) to first evaluate the transferability of our models. Later, an additional experiment using different channel sizes as input will be conducted to show the flexibility of the model for variation of inputs. To maintain consistency, all trials were down-sampled to 200Hz. Following a standard as the Shallow ConvNet benchmark \cite{BallCNN}, a 5th-order bandpass filter was applied between 4Hz and 32Hz, where motor imagery usually occurs. Normalisation across channels is done, followed by temporal normalisation across time steps to a zero mean and standard deviation of one. A batch size of 10 and a weight decay factor of 0.0005 were used for both \emph{shallowConv}-based Sandwich and \emph{EEGInception}-based Sandwich models. A learning rate of 0.001 and 0.0005 was used for shallowConv-Sandwich and EEGInception-Sandwich models respectively, aligning with their original studies. During the training of Sandwich models, four batches of size 10 from branches (40 in total with four branches) are delivered to the central layers simultaneously and features are distributed back to each separate local branch after common feature extraction. In the target training set, subject 1-3 has 100 trials each. Subjects 4 and 5 have 120 trials each. 20 trials each were used as the validation set for model selection. All randomisation and initialisation were conducted with a typical random seed 42 for machine learning for reproducibility in the same setup and environment. 

Given the disparate sample sizes in the three source sets (BCIC 2880 trials, Cho2017 9880 trials, PhysioMI 2399 trials), a potential imbalance in training individual feature extractors was anticipated. Since this study represents the initial attempts to test the Sandwich meta-framework systematically, efforts were made to minimise the impact of training imbalances during experimental design. While future studies may explore imbalanced sets to leverage full data sets, for the current investigation, trial numbers across all four data sets were balanced to 2880 trials, aligning with the size of BCIC sources. The PhysioMI and target data sets were augmented by duplicating the original data until reaching 2880 trials to address the varying sample sizes. The first 2880 trials of Cho2017 were utilised, and a random choice of samples was eschewed in favour of this method due to the deep set approach's requirement of subject IDs for grouping them into `sets'. These above processing procedures try to use as many data sets as possible but reduce the information of channels, trials or samples, whereas the winning method in the competition tried to maximise the channel sizes and window length to get higher performance but did not use all source sets. A comparison between the winning method and the Sandwich models under the same setup will be shown in the result section. 

\begin{figure*}
\centering
\includegraphics[scale=0.7]{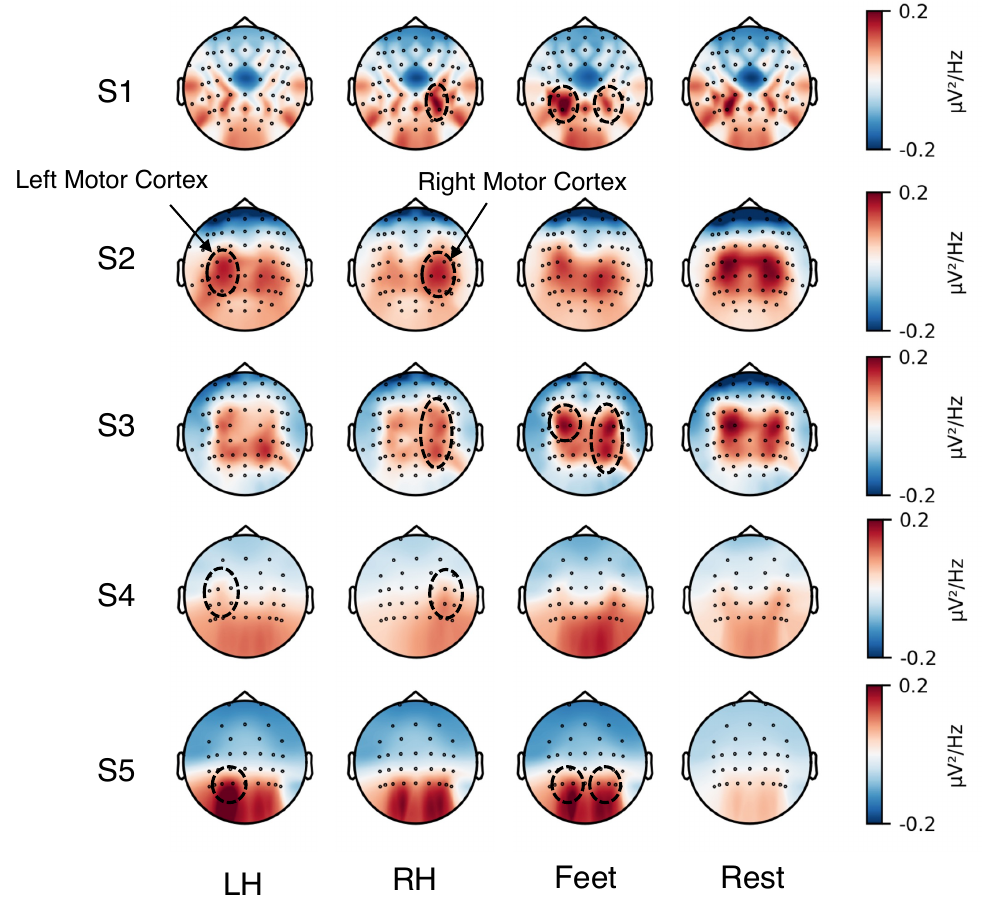}
\caption{Raw Data topography map of individual subjects in the test set}
\label{RawTopo}
\end{figure*}

As depicted in Figure \ref{RawTopo}, we analysed the topography showing the power map of channels for the five subjects within the test set to elucidate distinctive characteristics of EEG data during various tasks. To facilitate this visualisation, we leveraged the MNE topography API, as outlined in \cite{gramfort2013meg}. Each topomap has been normalised by subtracting the average power across all channels, thereby ensuring a uniform range of values from -0.2 to 0.2 for meaningful comparisons. Across subjects, a consistent pattern emerges, wherein a reduction in power is observed on the contralateral side of the motor cortex, specifically at channels C1, CP1, C2, and CP2, during the execution of left-hand and right-hand motor imagery tasks. When subjects engage in feet motor imagery, we note an increase in power around the motor cortex. This phenomenon could be explained by considering the reference channel, Cz, situated at the centre. It corresponds to the feet motor imagery task. Therefore, a decrease in power around the Cz channel could lead to an increase in power on both sides of the motor cortex, manifesting as the observed pattern during feet motor imagery.

\section{Results}
\begin{figure*}
\centering
\includegraphics[scale=0.8]{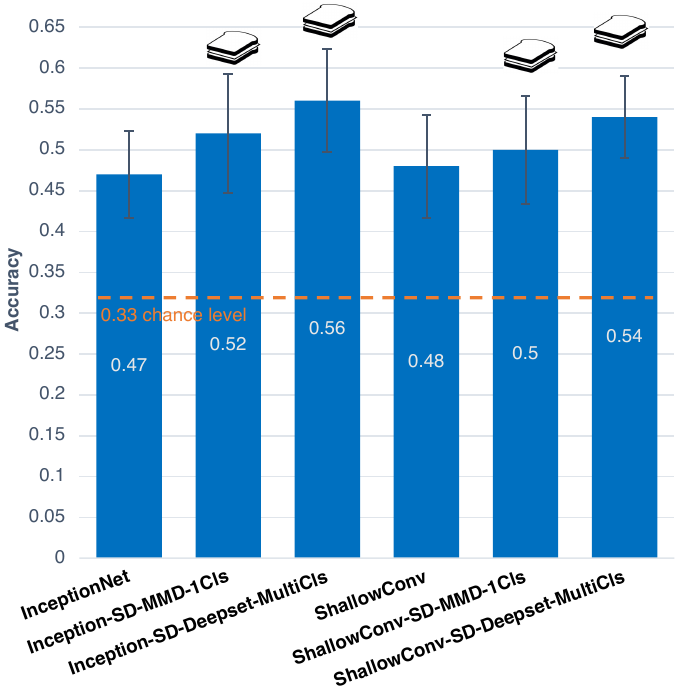}
\caption{Averaged decoding accuracy of different models}
\label{result_bar}
\end{figure*}

In this study, we compared two implementations of the Sandwich architectures, the ShallowConv-Sandwich (\emph{ShallowConv-SD}) models and the Inception-Sandwich (\emph{Inception-SD}) models, to the baseline \emph{Shallow ConvNet} and \emph{EEGInception}. Our approach encompasses two strategies addressing different labels across diverse tasks. The first involves a single classifier accommodating all potential labels within the data sets, while the second employs individual classifiers tailored to the specific labels of each dataset. We also tested two transfer learning approaches, the MMD and Deep Set, in the proposed framework. The decoding accuracy for various combinations of architectures and approaches is extensively presented in the appendix (see \ref{result_bar_all}). Within this section, we draw attention to the models that exhibit superior performance compared to the baseline models, as illustrated in Figure \ref{result_bar}.

The \emph{EEGInception} and \emph{Shallow ConvNet} exhibit comparable average decoding accuracies of 47\% and 48\%, establishing a baseline for comparison. The Single-classifier Sandwich models with the MMD alignment (\emph{SD-MMD-1Cls}) performed better than the baseline models, yielding a decoding accuracy of 52\% for the Inception-based model (\emph{Inception-SD-MMD-1Cls}), and 50\% shallowConvNet-based model (\emph{shallowConvNet-SD-MMD-1Cls}), as shown in Figure \ref{result_bar}. In terms of the performance of models with individual classifiers with deep-set layers (\emph{SD-Deepset-MultiCls}), Specifically, the Inception-based model (\emph{Inception-SD-Deepset-MultiCls}) exhibited a 9\% improvement over the \emph{EEGInception} baseline. A parallel trend was evident in the \emph{Shallow ConvNet} example, with \emph{ShallowConv-SD-Deepset-MultiCls} surpassing the \emph{shallowConv} net baseline by 6\%. 

Additional comparative models, including federated networks without transfer learning layers, networks combining multiple classifiers with MMD, and combining a single classifier with deep set layers, are detailed in the supplementary figure \ref{result_bar_all}. Compared to the baseline models, federated networks without MMD or deep set layers perform slightly better than the baseline, yielding decoding accuracy of 1\% and 3\% higher for the \emph{EEGinception} and \emph{Shallow ConvNet}. However, we can observe that in both cases, combining the deep set with a single classifier and combining MMD with multiple classifiers both result in lower performance. A clear example is that the \emph{Inception-SD-Deepset-1Cls} is 10\% lower than \emph{Inception-SD-Deepset-MultiCls}, and the \emph{Inception-SD-MMD-MultiCls} is 7\% lower than the performance of \emph{Inception-SD-MMD-1Cls}. We will provide insights into the rationale behind the superior performance of combining MMD with a single classification head and combining deep set with multiple classification heads in the Discussion section and explain why the transfer learning could not work by reserving the combination.

\begin{figure*}
\centering
\includegraphics[scale=0.7]{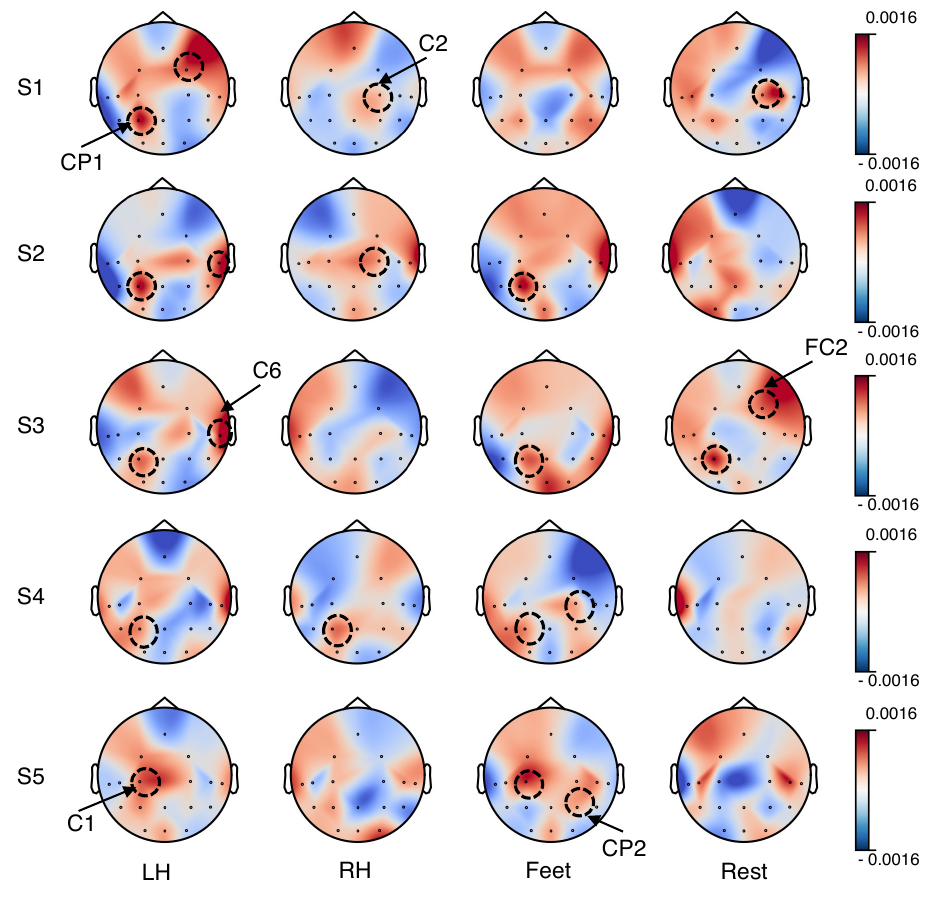}
\caption{Example Network Focus Visualisation on Inception-SD-Deepset-MultiCls}
\label{network_visualisation}
\end{figure*}

To provide interpretability for the model, we present a network-focused visualisation of tasks across five subjects in Figure \ref{network_visualisation}, utilising the \emph{Inception-Deepset-MultiCls} architecture with the highest decoding accuracy as an illustrative example. This visualisation is generated through an input-output perturbation method, as outlined in \cite{BallCNN}. The colour scheme employed reflects the relevance of data to the final prediction for each label, where red indicates a greater focus on specific regions. As depicted in the Figure, a discernible trend emerges, mirroring the topographical map of raw data around the motor cortex (Figure \ref{RawTopo}). Regions with relatively higher relevance, specifically C1, C2, CP1, and CP2, are circled around the motor cortex for the left hand (LH), right hand (RH), and feet across the five subjects. Additionally, the visualisation shows the algorithm sometimes tries to capture noisy channels during the learning process as features. Noteworthy instances include FP2 in the near frontal region for LH of subject 1, and the Rest state for subject 3, as well as C6 near the right ear for LH of subject 3. The discussion section will delve into the potential implications of motor-related features and noise-related features, offering a comprehensive exploration of the model's interpretability.



To elucidate the characteristics of the two proposed Sandwich architectures and provide empirical evidence of the impact of common transfer learning layers, we present the t-SNE \cite{van2008visualizing} plots comparing EEG features before and after the computation through the common layers. Specifically, we focus on the \emph{SD-MMD-1Cls} and \emph{SD-Deepset-MultiCls} models which showed superior performance than baselines, with \emph{ShallowConv-SD-MMD-1Cls} and \emph{ShallowConv-SD-Deepset-MultiCls} serving as illustrative examples. As articulated in Equation \ref{Equation2}, the MMD-based model aims to align different distributions of common labels (LH and RH) across data sets. Therefore, we visualise the learned features for LH and RH across four data sets, where shapes represent labels, and colors represent data sets. These t-SNE plots offer insights into how the common transfer learning layers influence the distributional alignment of features for the specified labels, providing a tangible representation of the impact of MMD-based alignment in the context of the Sandwich framework. 

\begin{figure*}
\centering
\includegraphics[scale=0.47]{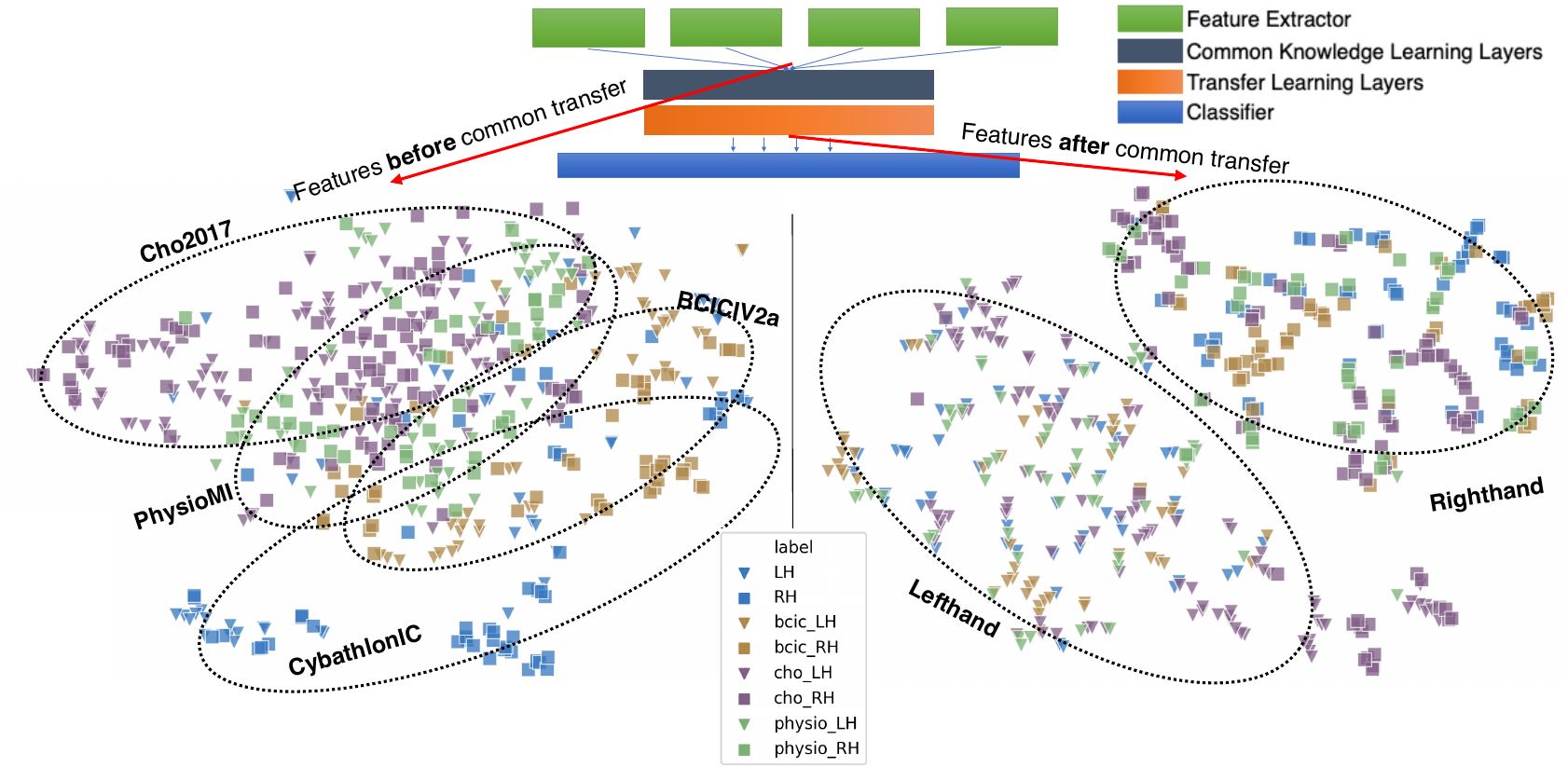}
\caption{t-SNE feature plot across data sets before (left) and after (right) the common transfer learning layers on \emph{ShallowConv-SD-MMD-1Cls}}
\label{tsne_MMD_1Cls}
\end{figure*}

As depicted in the t-SNE plot for \emph{ShallowConv-SD-MMD-1Cls} (Figure \ref{tsne_MMD_1Cls}), a distinct pattern emerges. Prior to the integration of common transfer learning layers, the four datasets exhibit separate clusters, indicating diverse distributions. However, after the application of common knowledge learning layers and MMD layers, two discernible effects come to light. One is that, post-alignment, the features from different data sets converge into relatively the same region, suggesting a harmonisation of distributions. This aligns with the goal of the MMD-based model to reduce the distributional gap among data sets. The other is the clustering of Left-Hand and Right-Hand labels. This outcome indicates that, despite the distributional alignment, the model has successfully retained the discriminative information between left-hand and right-hand labels, which is crucial for accurate classification. 

\begin{figure*}
\centering
\includegraphics[scale=0.46]{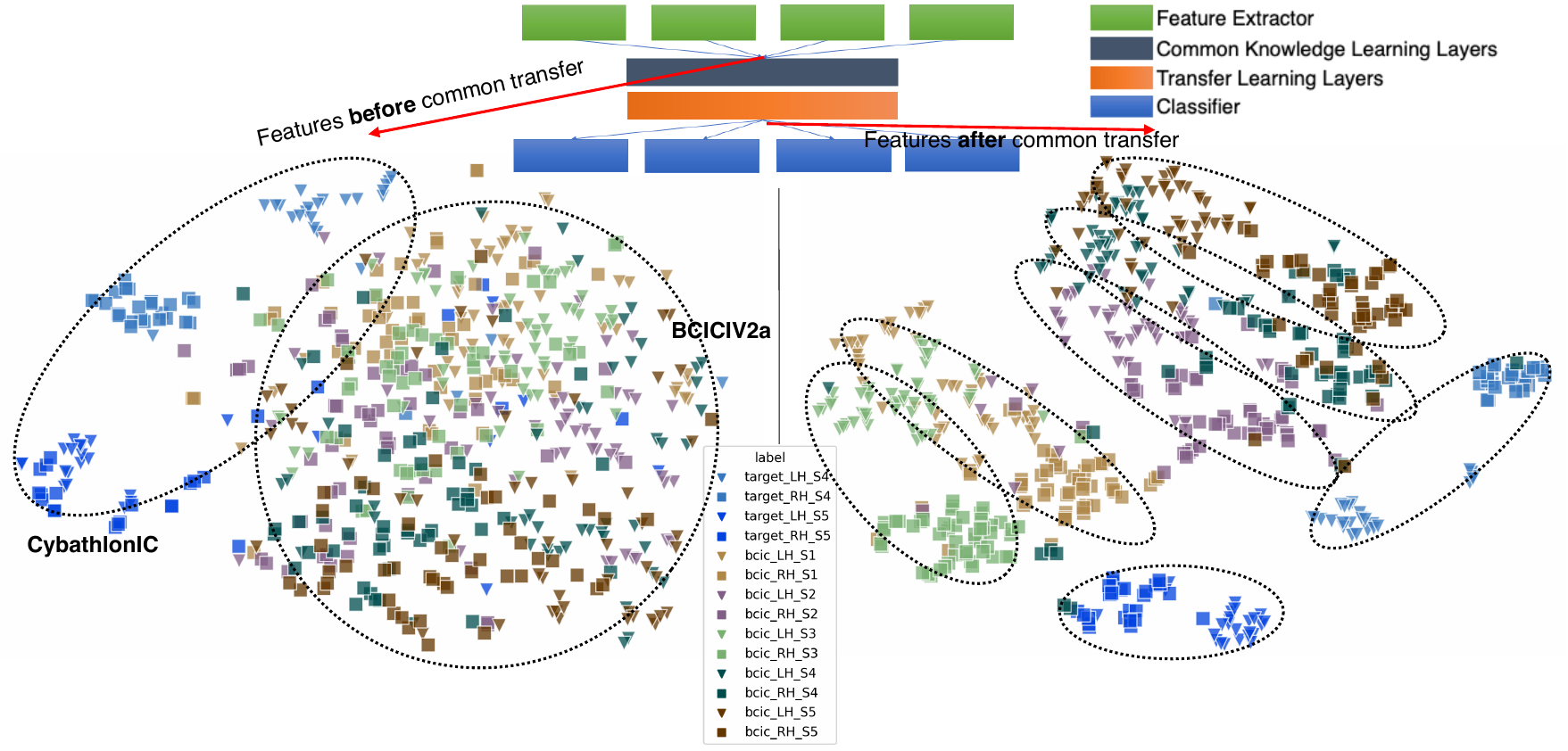}
\caption{t-SNE feature plot across data sets before (left) and after (right) the common transfer learning layers on \emph{ShallowConv-SD-Deepset-MultiCls} }
\label{tsne_Deepset_MultiCls}
\end{figure*}

As outlined in the methods and Theorem \ref{TheoremDeepset}, the Deepset-based model aims to enhance robustness and invariance within a set, considering subjects as sets in our context. To illustrate this effect, we present t-SNE plots focusing on the LH and RH labels for subjects within the CybathlonIC and BCICIV2a data sets. The shapes in the plot represent labels, while colours denote individual subjects. To manage visual complexity, we showcase the effect on seven subjects. As evident in Figure \ref{tsne_Deepset_MultiCls}, the t-SNE plots showcase the impact of the deepset-based model on the distribution of labels and subjects. Before the integration of common transfer layers, different labels and subjects did not exhibit clear clustering and classification patterns. However, post the application of common transfer layers, distinct clusters for different subjects become apparent. Moreover, different labels fall on different sides of these clusters, contributing to a clearer separation for classification purposes. The t-SNE plots for the inception-based models, presented in Appendix Figures \ref{tsne_MMD_1Cls_inception} and \ref{tsne_Deepset_MultiCls_inception}, exhibit a similar trend as above, further validating the effectiveness of the proposed approach across different model architectures.

The above experiments used all three source data sets and all tasks for demonstration of the ability of the `Sandwich' framework using heterogeneous data sets. To realise that purpose, we had to use merely 17 channels and a 3-second window which all data sets have in common. However, the winning benchmark in the BEETL Competition \cite{pmlr-v176-wei22a} only used the PhysioNet data set as the source set to use longer EEG trial (4-second window) and more common channels (30 instead of 17) to maintain more information for the test set, as well as using same tasks across target and sources. This approach results in a better baseline while decreasing the diversity and complexity of setups.  Additional approaches like label smoothing loss function were applied to increase the performance \cite{bakas2022team}. Although the benchmark algorithm reduced the complexity of the setup which limits the use of broader heterogeneous data sets, here we applied the Sandwich framework to the same setup as the benchmark in the competition as a comparison to the benchmark transfer learning algorithm in the literature. We reproduced the Benchmark \emph{EEGInception Deepset} model and got the same accuracy score as the benchmark in the competition (76\%), which details of the experiment could be found in their original study \cite{bakas2022team,pmlr-v176-wei22a}. This baseline was given by an additional average voting procedure during prediction, which trains ten models with different data splits and perform average voting in the end to get the best possible accuracy.

\begin{figure*}
\centering
\includegraphics[scale=0.6]{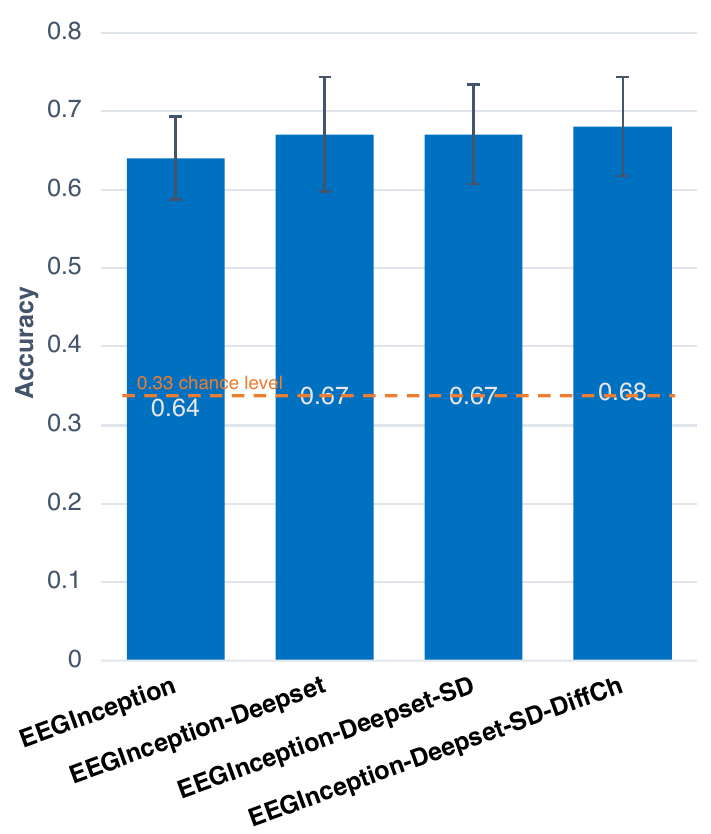}
\caption{Accuracies of models applying the Sandwich framework to the Benchmark Transfer Learning Model}
\label{SCSN_Deepset_Benchmark}
\end{figure*}

Figure \ref{SCSN_Deepset_Benchmark} shows the decoding accuracies of the benchmark model and the benchmark with the sandwich framework. In this comparison experiment, the main purpose is not to get the highest possible accuracy, here we show the comparison of the results of the models without the additional average voting approach. All transfer learning algorithms (benchmark and the sandwich networks) perform better than pure \emph{EEGInception} network baseline without transfer learning (64\%). The Sandwich network achieved comparable accuracy to the benchmark algorithm in the competition (67\%) while maintaining privacy-preserving properties through the protection of raw data and parameters, securely stored in the local branches. An additional experiment was conducted by varying channel sizes (\emph{DiffCh}), where the target set utilised 30 channels (consistent with the benchmark), and the source set employed all possible 64 channels, resulting in an accuracy of 68\%. This highlights the Sandwich network's flexibility in accommodating different input sizes within data sets, potentially expanding the utilisation of heterogeneous data sets, with additional privacy-preserving property.


\section{Discussion}
In the previous section, we conducted a performance comparison of the proposed Sandwich framework, highlighting the outcomes of employing various network backbones or transfer learning approaches. As demonstrated earlier, both Sandwich implementations exhibited superior performance compared to baseline methods across the example architectures. This observation underscores the potential of the Sandwich framework to effectively leverage cross-dataset features from diverse tasks, thereby enhancing the overall performance of EEG decoding. 

The Sandwich framework with multiple classifiers (\emph{SD-Deepset-MultiCls}) and the Sandwich frame with a unified classifier (\emph{SD-MMD-1Cls}) both outperformed the baseline with feasible transfer learning approaches. Here we highlight the commonalities and the differences between the two approaches. For both architectures, they used individual feature extractors which potentially handle differences in the raw data (e.g. impedance, sensor locations or channel sizes) by using different spatial and temporal kernels in the individual CNNs. Common knowledge learning layers are applied in both frameworks to learn the same set of network parameters (the same rules) to draw features from different data sets, which forces the network to learn the commonality across data sets. 

In both frameworks, it is evident that the application of appropriate transfer learning approaches can enhance the performance of models, whereas the requirements by the unified classifier and multiple classifiers are shown to be different. Both unified classifiers and multiple classifiers present potential solutions for handling label variability in data sets. However, the unified classifier necessitates that the features received for decoding adhere to a similar distribution. This requirement is essential for learning common discriminative rules for identical labels across different sets using the same classifier. Achieving this similarity in distribution mandates an alignment approach that projects features of the same labels from different data sets onto a comparable distribution. This alignment objective aligns with the design purpose of methods such as MMD, as evidenced by the t-SNE plot in Figure \ref{tsne_MMD_1Cls}. 

On the contrary, the Sandwich framework with multiple classifiers operates on a different premise. Unlike the unified classifier, it doesn't necessitate the labels of different data sets to conform to the same distribution. This is because individual data sets apply distinct decision boundaries for classification with individual classifiers.  Therefore, after the common knowledge learning layer, simply aligning data sets and features with the same label may not help the decoding, in fact, it has the potential to diminish discriminative information by projecting features to another latent space. In the supplementary experiment, when MMD alignment was applied to models with multiple classifiers, we observed a decrease in performance compared to models without MMD alignment. For instance, as depicted in supplementary Figure \ref{result_bar_all}, the accuracy of \emph{Inception-SD-MMD-MultiCls} (45\%) was lower than that without using MMD, such as \emph{Inception-SD-MultiCls} (48\%), and notably, 11\% lower than when using the Deepset approach, as seen in \emph{Inception-SD-Deepset-MultiCls}. The observed benefit of applying the deep set approach to multiple classifiers may be attributed to the impact of deep set layers on enhancing feature invariance within each subject considered as a `set'. This heightened feature invariance is reflected in the clustering of sets, as depicted in the t-SNE Figure \ref{tsne_Deepset_MultiCls}. Consequently, this increased feature invariance contributes to making the predictions of each classifier more robust. The deep set layers appear to play a crucial role in extracting and representing invariant features, fostering improved generalisation and performance of subjects within the multiple classifiers framework. 

The observed characteristics of the two frameworks prompt a discussion on the criteria for selecting between them. Briefly, the MMD approach with a single classifier aims to align distributions across samples to enhance classifier performance, yet it may diminish the unique dataset-specific features due to the unified projection. Conversely, the deep set framework employing multiple classifiers preserves subject-specific variability and individual discriminative information, although it does not guarantee perfect alignment across data sets, as evidenced by t-SNE plots. The challenge lies in the variance of the original sample distributions; the more divergent they are, the tougher it is to map features into a similar distribution. Consider sleep stage decoding and motor imagery as illustrative examples. Sleep stages are generally consistent across individuals, characterized by variations in brain energy levels, with most people adhering to this norm. In contrast, motor imagery involves more intricate functionalities of the motor cortex, where individual differences are more pronounced \cite{pmlr-v176-wei22a}. Hence, the decision between these approaches should hinge on the distributional disparities across data sets. For data sets with minor differences, an MMD-like method that projects them into a similar distribution could be beneficial. However, for more complex tasks that tap into intricate brain mechanisms, simple alignment might reduce each dataset's discriminative information. In such cases, a deep set framework with multiple classifiers could enhance invariance within sets while leveraging common knowledge layers for performance gains.

To gain insights into how a neural network perceives EEG data, a comparative analysis of the raw topography map in Figure \ref{RawTopo} and network relevance visualisation in Figure \ref{network_visualisation} across different tasks and subjects can be conducted. By examining these plots, inferences can be drawn about how the network's prediction rules align with our understanding of the human brain's biology. Common regions of interest, such as C1, C2, CP1, and CP2, emerge in both figures, aligning with the expected activation of the motor cortex during motor imagery tasks. However, there are still regions with high relevance that the network learnt that could be just noise of channels during sessions. For example, the high relevance of FP2 (near frontal region) for LH of subject 1 and Rest state for subject 3 could be just noise of eye movement or blinks that affect frontal channels. The high relevance of C6 near the right ear for LH of subject 3 could be ear or chewing movement during the experiment. Overfitting these noise features potentially decreases the generability of models on test sets. Based on these observations, further studies could explore methodologies for removing noisy features from deep learning models to enhance their generalisability. 

In addition to the benefit of enhancing decoding accuracies by utilising diverse data sets, we also emphasize four significant \textbf{privacy-preserving attributes} of the Sandwich framework. To begin with, subject-specific and dataset-specific information, i.e. raw data and IDs, is locally stored with the respective data owners, ensuring the preservation of \textbf{data-level privacy}. Moreover, the feature extractors are stored locally to maintain \textbf{parameter-level privacy}. In addition, local feature extractors naturally encrypt the raw data during feature extraction so as to \textbf{eliminate additional encryption costs} in traditional privacy-preserving methods. Lastly, labels and classifiers could be stored and predicted locally in a multi-classifier manner, safeguarding \textbf{inference-level privacy}. All these privacy-preserving properties are seamlessly integrated into the deep transfer learning architecture design.

We also illustrated the advantages of the Sandwich framework in comparison to the benchmark transfer learning algorithm. While exhibiting a comparable capacity to leverage the source set for improving model performance, reaching (67\%) as depicted in Figure \ref{SCSN_Deepset_Benchmark}, the Sandwich network demonstrated two additional key advantages: enhanced privacy preservation and the capability to accommodate diverse input shapes. This adaptability facilitates the utilisation of heterogeneous data sets with varying configurations, as the `\emph{DiffCh}' experiment in Figure \ref{SCSN_Deepset_Benchmark}, which extends the Deep set sandwich to the different input channel sizes, with a comparable accuracy of 68\%. The ability to adapt to diverse input shapes holds the potential to extend the applicability of algorithms in the literature, enabling their utilisation with a broader range of heterogeneous data sets. 

Moreover, the Sandwich framework represents an architecture-agnostic meta-framework with the potential for application in domains beyond motor imagery EEG decoding. As described in the methodology section and implied in the results, the feature extractors and task decoding layers can be readily substituted with different models, e.g. shallow CNNs, Inception networks, and potentially other task-specific decoders. These layers are flexible to be replaced with alternative time-series decoding layers, such as EMG decoding networks or domain-specific layers for ECoG, fMRI, etc. This flexibility highlights the versatility of the Sandwich meta-architecture, enabling its adaptation to various domains and tasks beyond EEG decoding.

It's important to highlight certain limitations and prerequisites associated with the use of the Sandwich framework. This approach, involving split individual feature extractors, is particularly beneficial when there exists a substantial difference in the distributions of various data sets, for example, in motor imagery tasks. In scenarios where the distributional shift is relatively consistent, as in sleep stage decoding, opting for a common feature extractor may be a more advantageous design \cite{pmlr-v176-wei22a}. This is because a shared feature extractor allows the model to learn and adapt to small shifts in distribution autonomously. However, when individual feature extractors are split, it can potentially reduce the amount of data available to each branch, leading to overfitting and diminishing generalisability, especially if the distributional shift is not pronounced. Moreover, it's crucial to consider the size gap among the target and source sets. In this study, there are several hundred examples in the target set and several thousand in each source set. It's worth noting that if the over-sampling of the target set is over-conducted to compensate for data shortage, the target feature extractor may overfit quickly compared to other feature extractors. This can result in unbalanced training, potentially undermining the effectiveness of transfer learning and alignment approaches. Therefore, careful consideration of data set characteristics and sizes is essential when applying the Sandwich framework to ensure optimal model performance. In addition, synchronisation during training \cite{baptista2012collective} is another engineering problem that needs to be considered. In this study, we examine the theoretical applicability of the model. However, in reality, how to achieve asynchronous communication and training is a technical problem that needs to be solved. Furthermore, the generalisation of our proposed models across broader brainwave datasets other than EEG poses a future direction. There also could be certain properties of other data types that EEG could not cover, fMRI, for example, is more similar to images instead of time series \cite{glover2011overview}, in which the feasibility of our method needs to be tested and refined according to such specific domain when applied.

In conclusion, we introduced the Sandwich meta-architecture, exemplified through the \emph{Inception-SD} and \emph{ShallowConv-SD} models with transfer learning approaches. These implementations showcase a federated deep transfer learning technique designed for cross-data sets and cross-task applications. Our approach integrates privacy-preserving features with deep transfer learning methodologies, leveraging transfer learning approaches to enhance model performance. The results demonstrate that the proposed Sandwich framework, benefiting from both transfer learning and privacy-preserving characteristics, outperforms the baseline methods. This suggests that our proposed methods have the potential to effectively leverage larger, heterogeneous data sets encompassing diverse tasks for transfer learning. Notably, the Sandwich meta-framework exhibits improved privacy-preserving properties at the raw data level, parameter level and inference level. This indicates a promising avenue for advancing the application of deep transfer learning techniques in real-life scenarios with enhanced privacy considerations across borders. Furthermore, our findings suggest that the Sandwich framework holds promise as an architecture-agnostic meta-framework, capable of integrating decoding architectures and transfer learning approaches for time series data. This versatility opens up possibilities for broader applications and encourages exploration into how diverse decoding architectures could be incorporated within the Sandwich framework to improve the decoding of EEG and time series data.

\noindent
\\
\textbf{References}
\\

\bibliographystyle{vancouver}
\bibliography{reference}

\begin{thebibliography}{10}

\bibitem{vaid2015eeg}
Vaid S, Singh P, Kaur C.
\newblock EEG signal analysis for BCI interface: A review.
\newblock In: 2015 fifth international conference on advanced computing \& communication technologies. IEEE; 2015. p. 143-7.

\bibitem{rashid2020current}
Rashid M, Sulaiman N, PP~Abdul~Majeed A, Musa RM, Ab~Nasir AF, Bari BS, et~al.
\newblock Current status, challenges, and possible solutions of EEG-based brain-computer interface: a comprehensive review.
\newblock Frontiers in neurorobotics. 2020;14:515104.

\bibitem{ramadan2017brain}
Ramadan RA, Vasilakos AV.
\newblock Brain computer interface: control signals review.
\newblock Neurocomputing. 2017;223:26-44.

\bibitem{walker2015deep}
Walker I, Deisenroth M, Faisal A.
\newblock Deep convolutional neural networks for brain computer interface using motor imagery.
\newblock Imperial College, Tech Report. 2015:68.

\bibitem{BallCNN}
Schirrmeister RT, Springenberg JT, Fiederer LDJ, Glasstetter M, Eggensperger K, Tangermann M, et~al.
\newblock Deep learning with convolutional neural networks for EEG decoding and visualization.
\newblock Hum Brain Mapp. 2017;38(11):5391-420.

\bibitem{4}
Acharya UR, Oh SL, Hagiwara Y, Tan JH, Adeli H.
\newblock Deep convolutional neural network for the automated detection and diagnosis of seizure using EEG signals.
\newblock Computers in biology and medicine. 2018;100:270-8.

\bibitem{5}
Oh SL, Hagiwara Y, Raghavendra U, Yuvaraj R, Arunkumar N, Murugappan M, et~al.
\newblock A deep learning approach for Parkinson’s disease diagnosis from EEG signals.
\newblock Neural Computing and Applications. 2018:1-7.

\bibitem{6}
Dai M, Zheng D, Na R, Wang S, Zhang S.
\newblock EEG classification of motor imagery using a novel deep learning framework.
\newblock Sensors. 2019;19(3):551.

\bibitem{3}
Schirrmeister RT, Springenberg JT, Fiederer LDJ, Glasstetter M, Eggensperger K, Tangermann M, et~al.
\newblock Deep learning with convolutional neural networks for EEG decoding and visualization.
\newblock Human brain mapping. 2017;38(11):5391-420.

\bibitem{23}
Ortega P, Colas C, Faisal AA.
\newblock Compact convolutional neural networks for multi-class, personalised, closed-loop EEG-BCI.
\newblock In: 2018 7th IEEE International Conference on Biomedical Robotics and Biomechatronics (Biorob). IEEE; 2018. p. 136-41.

\bibitem{24}
Tayeb Z, Fedjaev J, Ghaboosi N, Richter C, Everding L, Qu X, et~al.
\newblock Validating deep neural networks for online decoding of motor imagery movements from EEG signals.
\newblock Sensors. 2019;19(1):210.

\bibitem{25}
Pan SJ, Yang Q.
\newblock A survey on transfer learning.
\newblock IEEE Transactions on knowledge and data engineering. 2009;22(10):1345-59.

\bibitem{wei2021inter}
Wei X, Ortega P, Faisal AA.
\newblock Inter-subject deep transfer learning for motor imagery eeg decoding.
\newblock In: 2021 10th International IEEE/EMBS Conference on Neural Engineering (NER). IEEE; 2021. p. 21-4.

\bibitem{weiss2016survey}
Weiss K, Khoshgoftaar TM, Wang D.
\newblock A survey of transfer learning.
\newblock Journal of Big data. 2016;3(1):1-40.

\bibitem{11}
Goldberger AL, et~al.
\newblock PhysioBank, PhysioToolkit, and PhysioNet: components of a new research resource for complex physiologic signals.
\newblock circulation. 2000;101(23):e215-20.

\bibitem{12}
Huang JT, Li J, Yu D, Deng L, Gong Y.
\newblock Cross-language knowledge transfer using multilingual deep neural network with shared hidden layers.
\newblock In: 2013 IEEE International Conference on Acoustics, Speech and Signal Processing. IEEE; 2013. p. 7304-8.

\bibitem{13}
Long M, Cao Y, Wang J, Jordan MI.
\newblock Learning transferable features with deep adaptation networks.
\newblock arXiv preprint arXiv:150202791. 2015.

\bibitem{jayaram2016transfer}
Jayaram V, Alamgir M, Altun Y, Scholkopf B, Grosse-Wentrup M.
\newblock Transfer learning in brain-computer interfaces.
\newblock IEEE Computational Intelligence Magazine. 2016;11(1):20-31.

\bibitem{lotte2018review}
Lotte F, Bougrain L, Cichocki A, Clerc M, Congedo M, Rakotomamonjy A, et~al.
\newblock A review of classification algorithms for EEG-based brain--computer interfaces: a 10 year update.
\newblock Journal of neural engineering. 2018;15(3):031005.

\bibitem{14}
Tan C, Sun F, Zhang W.
\newblock Deep transfer learning for EEG-based brain computer interface.
\newblock In: 2018 IEEE International Conference on Acoustics, Speech and Signal Processing (ICASSP). IEEE; 2018. p. 916-20.

\bibitem{7}
Tan C, Sun F, Fang B, Kong T, Zhang W.
\newblock Autoencoder-based transfer learning in brain--computer interface for rehabilitation robot.
\newblock International Journal Of Advanced Robotic Systems. 2019;16(2):1729881419840860.

\bibitem{15}
Sakhavi S, Guan C.
\newblock Convolutional neural network-based transfer learning and knowledge distillation using multi-subject data in motor imagery BCI.
\newblock In: 2017 8th International IEEE/EMBS Conference on Neural Engineering (NER). IEEE; 2017. p. 588-91.

\bibitem{17}
Hang W, Feng W, Du R, Liang S, Chen Y, Wang Q, et~al.
\newblock Cross-subject EEG signal recognition using deep domain Adaptation network.
\newblock IEEE Access. 2019;7:128273-82.

\bibitem{pmlr-v176-wei22a}
Wei X, et~al.
\newblock 2021 BEETL Competition: Advancing Transfer Learning for Subject Independence \& Heterogenous EEG Data Sets.
\newblock In: Proceedings of the NeurIPS 2021 Competitions and Demonstrations Track. vol. 176 of Proceedings of Machine Learning Research. PMLR; 2022. p. 205-19.

\bibitem{smola2006maximum}
Smola AJ, Gretton A, Borgwardt K.
\newblock Maximum mean discrepancy.
\newblock In: 13th international conference, ICONIP; 2006. p. 3-6.

\bibitem{li2020maximum}
Li J, Chen E, Ding Z, Zhu L, Lu K, Shen HT.
\newblock Maximum density divergence for domain adaptation.
\newblock IEEE transactions on pattern analysis and machine intelligence. 2020;43(11):3918-30.

\bibitem{han2023eeg}
Han J, Wei X, Faisal AA.
\newblock EEG Decoding for Datasets with Heterogenous Electrode Configurations using Transfer Learning Graph Neural Networks.
\newblock arXiv preprint arXiv:230613109. 2023.

\bibitem{zaheer2017deep}
Zaheer M, Kottur S, Ravanbakhsh S, Poczos B, Salakhutdinov RR, Smola AJ.
\newblock Deep sets.
\newblock Advances in neural information processing systems. 2017;30.

\bibitem{bakas2022team}
Bakas S, Ludwig S, Barmpas K, Bahri M, Panagakis Y, Laskaris N, et~al.
\newblock Team cogitat at neurips 2021: Benchmarks for eeg transfer learning competition.
\newblock arXiv preprint arXiv:220203267. 2022.

\bibitem{li2020review}
Li L, et~al.
\newblock A review of applications in federated learning.
\newblock Computers \& Industrial Engineering. 2020;149:106854.

\bibitem{hao2019towards}
Hao M, Li H, Xu G, Liu S, Yang H.
\newblock Towards efficient and privacy-preserving federated deep learning.
\newblock In: 2019 IEEE international conference on communications (ICC). IEEE; 2019. p. 1-6.

\bibitem{lyu2017privacy}
Lyu L, He X, Law YW, Palaniswami M.
\newblock Privacy-preserving collaborative deep learning with application to human activity recognition.
\newblock In: Proceedings of the 2017 ACM on Conference on Information and Knowledge Management; 2017. p. 1219-28.

\bibitem{dong2017dropping}
Dong H, Wu C, Wei Z, Guo Y.
\newblock Dropping activation outputs with localized first-layer deep network for enhancing user privacy and data security.
\newblock IEEE Transactions on Information Forensics and Security. 2017;13(3):662-70.

\bibitem{du2001secure}
Du W, Atallah MJ.
\newblock Secure multi-party computation problems and their applications: a review and open problems.
\newblock In: Proceedings of the 2001 workshop on New security paradigms; 2001. p. 13-22.

\bibitem{truong2021privacy}
Truong N, Sun K, Wang S, Guitton F, Guo Y.
\newblock Privacy preservation in federated learning: An insightful survey from the GDPR perspective.
\newblock Computers \& Security. 2021;110:102402.

\bibitem{popescu2021privacy}
Popescu AB, et~al.
\newblock Privacy preserving classification of eeg data using machine learning and homomorphic encryption.
\newblock Applied Sciences. 2021;11(16):7360.

\bibitem{xia2022privacy}
Xia K, Duch W, Sun Y, Xu K, Fang W, Luo H, et~al.
\newblock Privacy-Preserving Brain--Computer Interfaces: A Systematic Review.
\newblock IEEE Trans Comput Soc Syst. 2022.

\bibitem{ju2020federated}
Ju C, Gao D, Mane R, Tan B, Liu Y, Guan C.
\newblock Federated transfer learning for EEG signal classification.
\newblock In: 2020 42nd Conf Proc IEEE Eng Med Biol Soc (EMBC). IEEE; 2020. p. 3040-5.

\bibitem{bethge2022domain}
Bethge D, Hallgarten P, Grosse-Puppendahl T, Kari M, Mikut R, Schmidt A, et~al.
\newblock Domain-Invariant Representation Learning from EEG with Private Encoders.
\newblock In: ICASSP 2022. IEEE; 2022. p. 1236-40.

\bibitem{kapitonova2022framework}
Kapitonova M, Kellmeyer P, Vogt S, Ball T.
\newblock A Framework for Preserving Privacy and Cybersecurity in Brain-Computer Interfacing Applications.
\newblock arXiv preprint arXiv:220909653. 2022.

\bibitem{santamaria2020eeg}
Santamaria-Vazquez E, Martinez-Cagigal V, Vaquerizo-Villar F, Hornero R.
\newblock EEG-inception: A novel deep convolutional neural network for assistive ERP-based brain-computer interfaces.
\newblock IEEE Trans Neural Syst Rehabil Eng. 2020;28(12):2773-82.

\bibitem{long2015learning}
Long M, Cao Y, Wang J, Jordan M.
\newblock Learning transferable features with deep adaptation networks.
\newblock In: International conference on machine learning. PMLR; 2015. p. 97-105.

\bibitem{bakas2023latent}
Bakas S, Ludwig S, Adamos DA, Laskaris N, Panagakis Y, Zafeiriou S.
\newblock Latent Alignment with Deep Set EEG Decoders.
\newblock arXiv preprint arXiv:231117968. 2023.

\bibitem{Cho2017}
Cho H, et~al.
\newblock {EEG} datasets for motor imagery brain--computer interface.
\newblock GigaScience. 2017;6(7):gix034.

\bibitem{BCIC}
Tangermann M, M{\"u}ller KR, Aertsen A, Birbaumer N, Braun C, Brunner C, et~al.
\newblock Review of the BCI competition IV.
\newblock Front Neurosci. 2012;6:55.

\bibitem{hwaidi2022classification}
Hwaidi JF, Chen TM.
\newblock Classification of motor imagery EEG signals based on deep autoencoder and convolutional neural network approach.
\newblock IEEE access. 2022;10:48071-81.

\bibitem{schalk2004bci2000}
Schalk G, et~al.
\newblock BCI2000: a general-purpose brain-computer interface (BCI) system.
\newblock IEEE Trans Biomed Eng. 2004;51(6):1034-43.

\bibitem{weibo}
Yi W, Qiu S, Wang K, Qi H, Zhang L, Zhou P, et~al.
\newblock Evaluation of EEG oscillatory patterns and cognitive process during simple and compound limb motor imagery.
\newblock PloS one. 2014;9(12):e114853.

\bibitem{gramfort2013meg}
Gramfort A, Luessi M, Larson E, Engemann DA, Strohmeier D, Brodbeck C, et~al.
\newblock MEG and EEG data analysis with MNE-Python.
\newblock Frontiers in neuroscience. 2013:267.

\bibitem{van2008visualizing}
Van~der Maaten L, Hinton G.
\newblock Visualizing data using t-SNE.
\newblock Journal of machine learning research. 2008;9(11).

\bibitem{baptista2012collective}
Baptista MS, Ren HP, Swarts JC, Carareto R, Nijmeijer H, Grebogi C.
\newblock Collective almost synchronisation in complex networks.
\newblock PloS one. 2012;7(11):e48118.

\bibitem{glover2011overview}
Glover GH.
\newblock Overview of functional magnetic resonance imaging.
\newblock Neurosurgery Clinics. 2011;22(2):133-9.

\end{thebibliography}

\appendix
\section{}

\begin{table}[H]
\centering
\caption{MI data sets in BEETL}
\begin{tabular}{llll}
\hline
MI Data set             & Subjects & Channels & Tasks                                \\ \hline
Cho2017               & 52       & 64       & Left/Right hand                 \\
BNCI2014               & 9        & 22       & Left/Right hand/Feet/Tongue     \\
PhysionetMI            & 109      & 64       & Left/Right hand/Feet/Both hands/Rest \\ 
Weibo2014                & 10       & 60       & Left/Right hand/Feet/Rest       \\
CybathlonIC & 5        & 63       & Left/Right hand/Feet/Rest \\ \hline
\end{tabular}
\label{MIdataset}
\end{table}

\begin{figure}[H]
\centering
\includegraphics[scale=0.42]{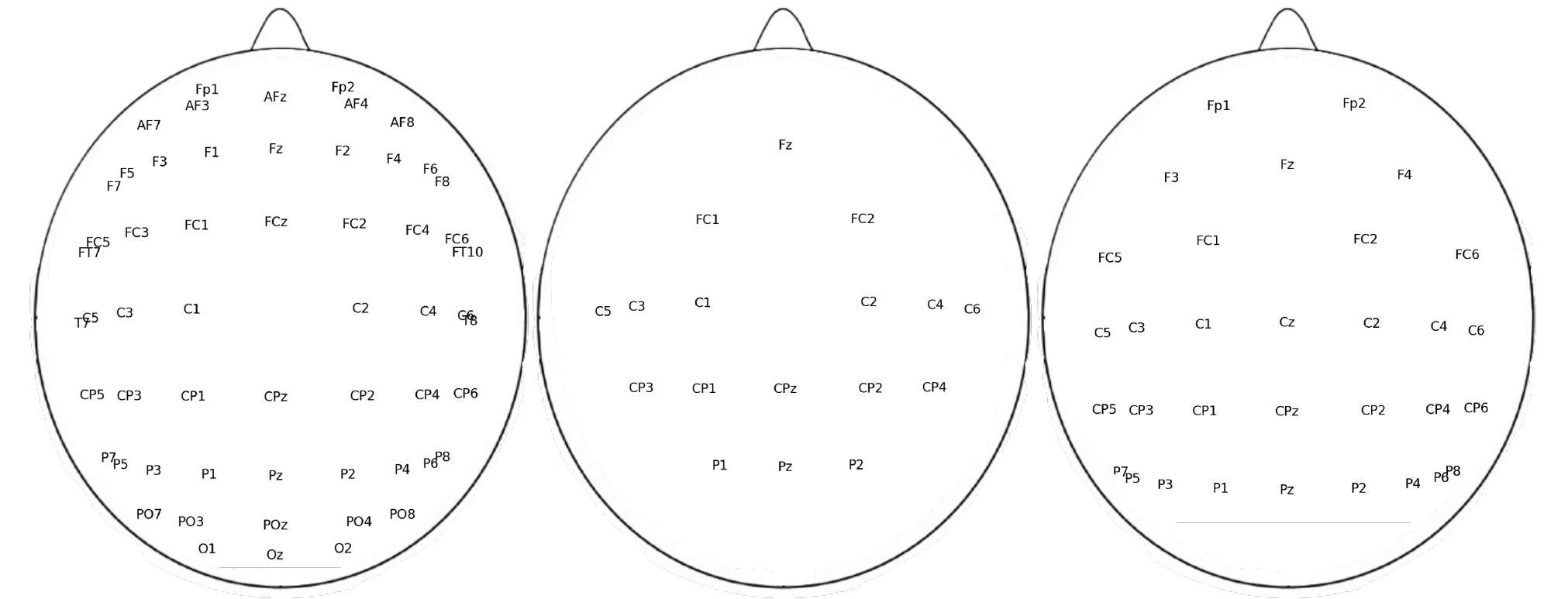}
\caption{Channel positions of test set A (Left), test set B (Right) and their common channels (Middle)}
\label{ChannelPositions}
\end{figure}

\begin{figure}[H]
\centering
\includegraphics[scale=0.5]{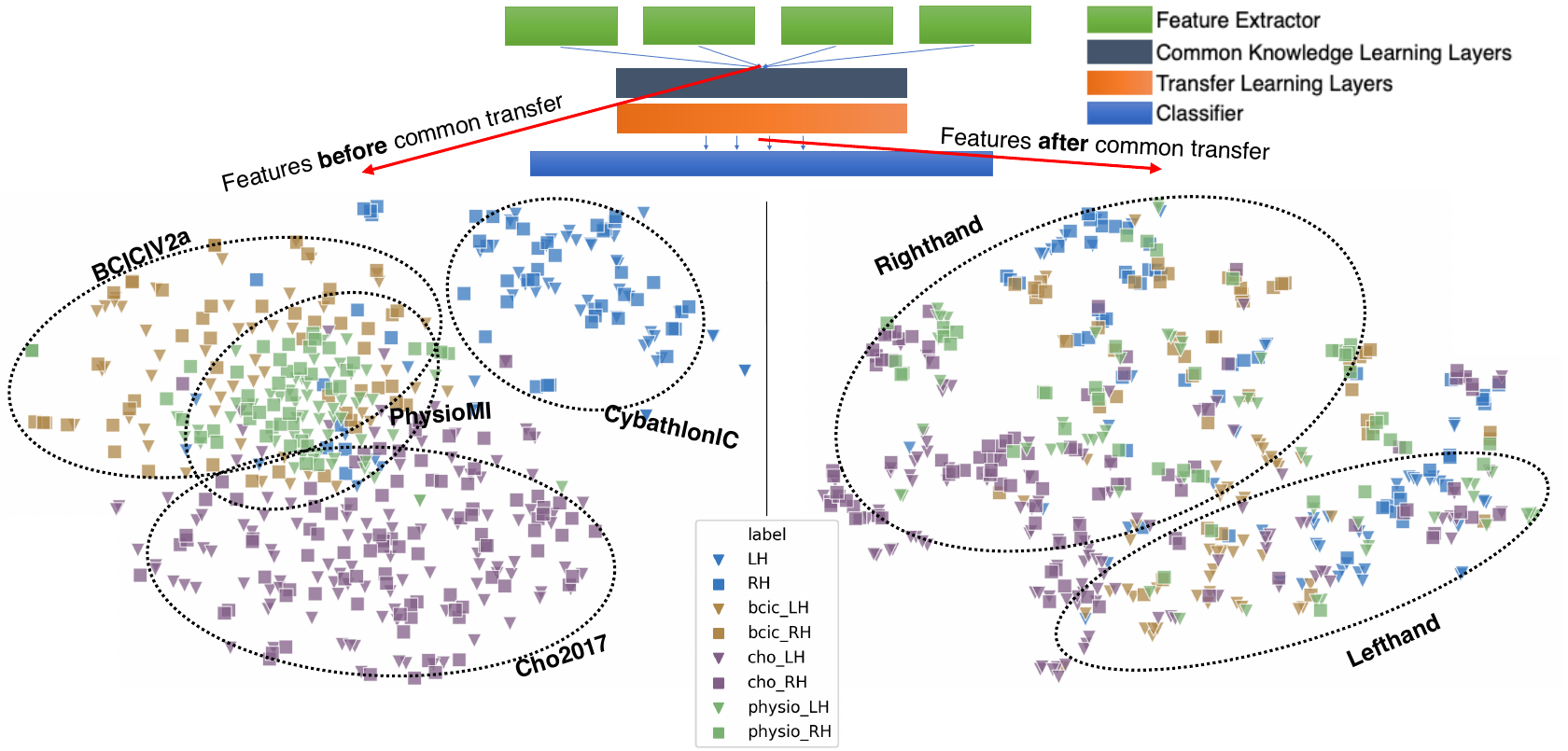}
\caption{t-SNE feature plot across data sets before (left) and after (right) the common transfer learning layers on \emph{Inception-SD-MMD-1Cls}}
\label{tsne_MMD_1Cls_inception}
\end{figure}

\begin{figure}[H]
\centering
\includegraphics[scale=0.5]{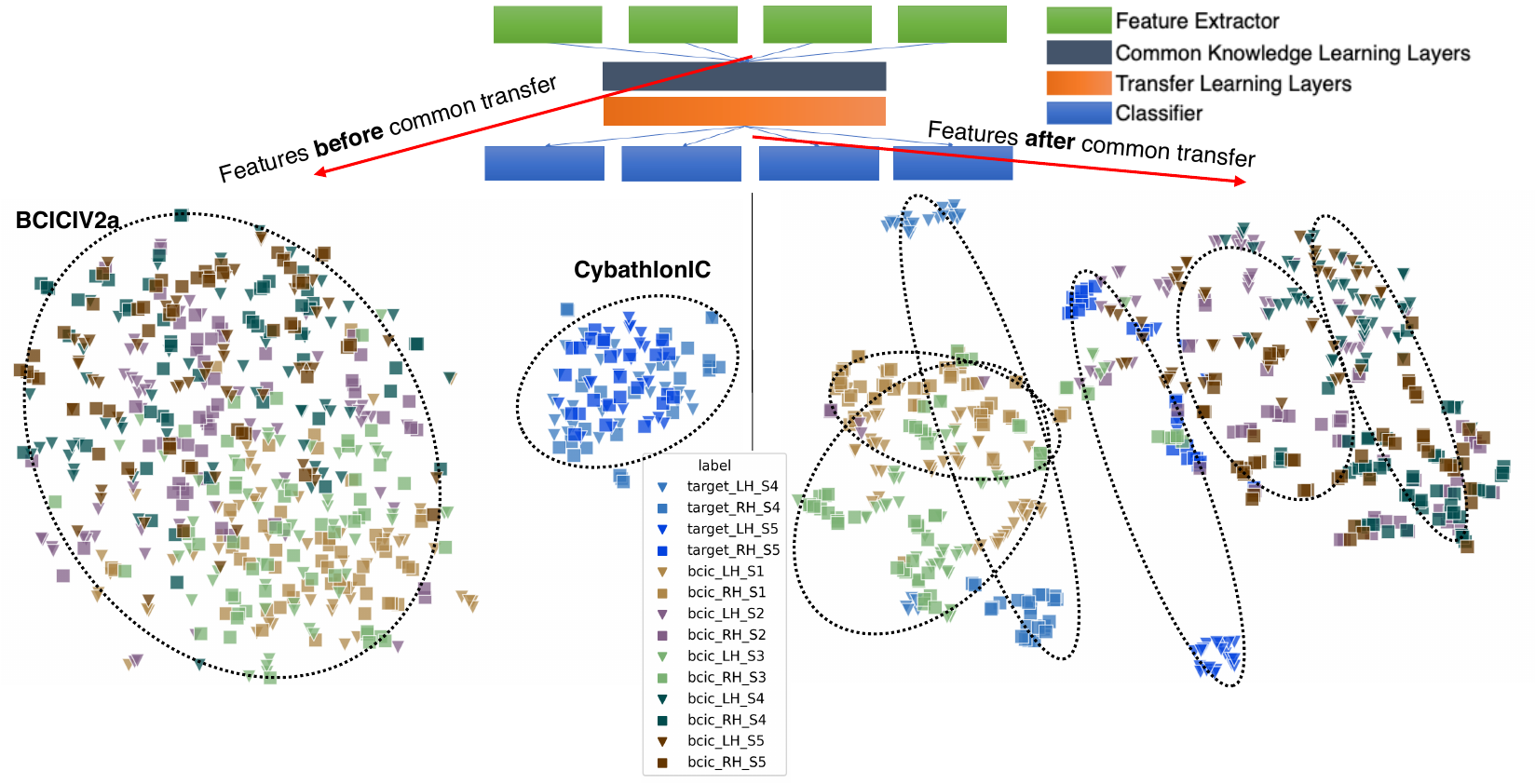}
\caption{t-SNE feature plot across data sets before (left) and after (right) the common transfer learning layers on \emph{Inception-SD-Deepset-MultiCls}}
\label{tsne_Deepset_MultiCls_inception}
\end{figure}

\begin{figure}[H]
\centering
\includegraphics[scale=0.6]{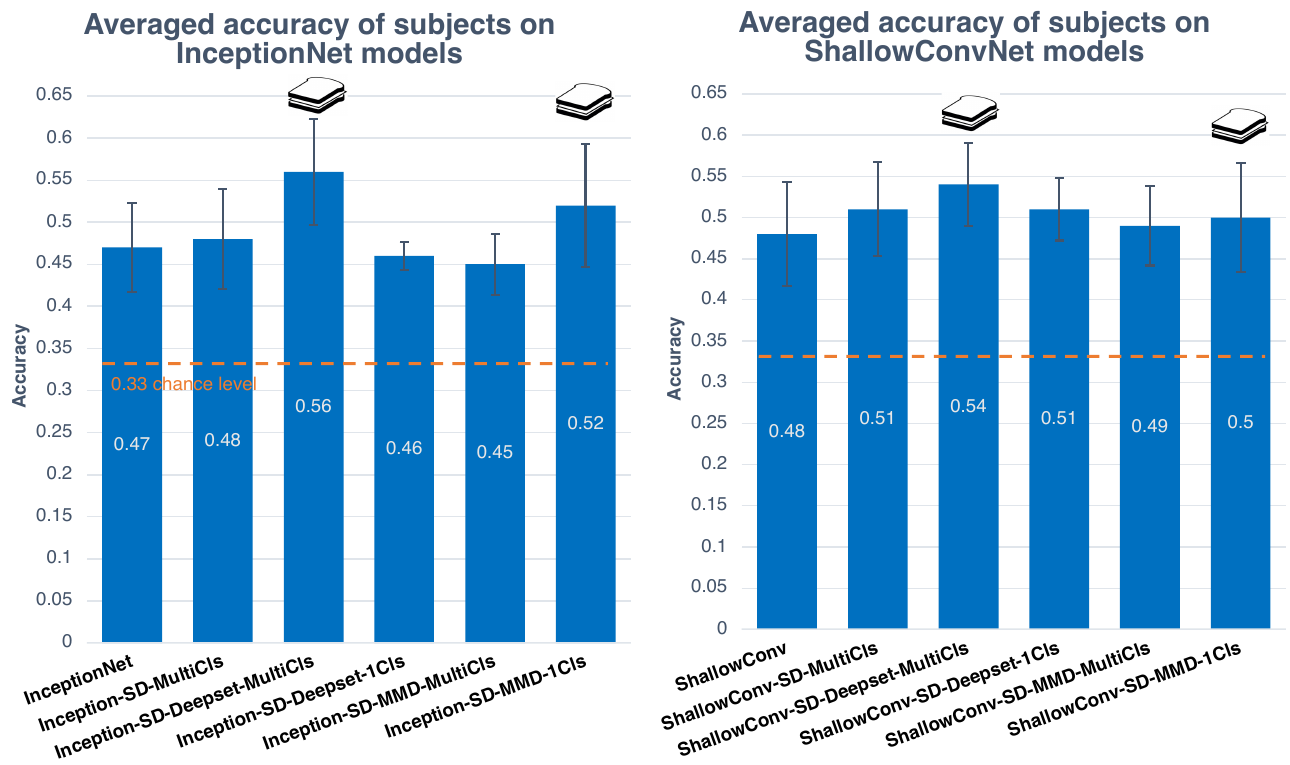}
\caption{Averaged decoding accuracy of models with the combination of approaches }
\label{result_bar_all}
\end{figure}

\end{document}